\documentclass{aastex}
\usepackage{emulateapj5}

\begin{document}

\title{A Simple Model for the Absorption of Starlight by
Dust in Galaxies}
\author{St\'ephane Charlot\altaffilmark{1}}
\affil{Institut d'Astrophysique de Paris, CNRS, 98 bis Boulevard Arago, 75014 Paris, France;
charlot@iap.fr}
\email{charlot@iap.fr}
\and
\author{S. Michael Fall}
\affil{Space Telescope Science Institute, 3700 San Martin Drive, Baltimore, MD 21218;
fall@stsci.edu}
\email{fall@stsci.edu}

\altaffiltext{1}{Also Max-Planck Institut f\"ur Astrophysik, Karl-Schwarzschild-Strasse 1,
85748 Garching, Germany}

\submitted{Received 1999 July 29 ; accepted 2000 February 28}

\begin{abstract}
\small
We present a new model to compute the effects of dust on the 
integrated spectral properties of galaxies, based on an
idealized prescription of the main features of the interstellar
medium (ISM). The model includes the ionization of H{\sc ii}
regions in the interiors of the dense clouds in which stars form 
and the influence of the finite lifetime of these clouds on the
absorption of radiation. We compute the production of emission
lines and the absorption of continuum radiation in the H{\sc ii} 
regions and the subsequent transfer of line and continuum radiation
in the surrounding H{\sc i} regions and the ambient ISM. This 
enables us to interpret simultaneously all the observations
of a homogeneous sample of nearby ultraviolet-selected starburst
galaxies, including the
ratio of far-infrared to ultraviolet luminosities, the ratio of 
H$\alpha$ to H$\beta$ luminosities, the H$\alpha$ equivalent width,
and the ultraviolet spectral slope. We show that the finite 
lifetime of stellar birth clouds is a key ingredient to
resolve an apparent discrepancy between the attenuation of 
line and continuum photons in starburst galaxies. In addition, 
we find that an effective absorption curve proportional to 
$\lambda^{ -0.7}$ reproduces the observed relation between the 
ratio of far-infrared to ultraviolet luminosities and the ultraviolet 
spectral slope. We interpret this relation most simply as a
sequence in the overall dust content of the galaxies. The shallow
wavelength dependence of the effective absorption curve is 
compatible with the steepness of known extinction curves 
if the dust has a patchy distribution. In particular, we find
that a random distribution of discrete clouds with optical 
depths similar to those in the Milky Way provides a consistent
interpretation of all the observations. A noteworthy outcome 
of our detailed analysis is that the observed mean relations 
for starburst galaxies can be closely approximated by the 
following simple recipe: use an effective absorption curve 
proportional to $\lambda^{-0.7}$ to attenuate the line and 
continuum radiation from each stellar generation, and lower 
the normalization of the curve typically by a factor of 3 
after $10^7\,$yr to account for the dispersal of the birth 
clouds. This recipe or our full model for absorption can be 
incorporated easily into any population synthesis model.
\end{abstract}

\keywords{galaxies: ISM --- galaxies: starburst --- infrared: 
galaxies --- ISM: dust, extinction --- ultraviolet: galaxies}

\section{Introduction}

To interpret the observed spectral properties of galaxies,
we require models for both the production of stellar radiation 
and its transfer through the interstellar medium (ISM). Currently,
the accuracy of population synthesis models contrasts with the
rudimentary way in which the absorption of starlight by dust is
often treated. In many applications, dust is either
ignored or assumed to be distributed in a uniform screen in 
front of the stars. The resulting uncertainties in the 
absorption of the ultraviolet radiation in galaxies can be as 
much as an order of magnitude. This problem has become 
especially acute in studies of galaxies at high redshifts. 
Nearby starburst galaxies may be suitable analogs of 
high-redshift galaxies and provide important clues to 
interpreting their spectral properties. The observations of
nearby ultraviolet-selected starburst galaxies are numerous, 
including the ratio of far-infrared to ultraviolet luminosities, 
the ratio of H$\alpha$ to H$\beta$ luminosities, the H$\alpha$
equivalent width, and the ultraviolet spectral slope. In 
particular, there is a remarkably tight correlation between
far-infrared luminosity and ultraviolet spectral slope (Meurer et
al. 1995; Meurer, Heckman, \& Calzetti 1999). This wealth of 
observations can potentially help us quantify the effects of
dust on various spectral properties of galaxies.

There have been several analyses of the spatial distribution
and optical properties of the dust in nearby starburst galaxies 
based on various subsets of the observations (e.g., Fanelli, 
O'Connell, \& Thuan 1988; Calzetti, Kinney, \& Storchi-Bergmann
1994, 1996; Puxley \& Brand 1994; Meurer et al. 1995; Gordon, 
Calzetti, \& Witt 1997). A generic result of these studies is 
that if the dust is distributed in a uniform foreground screen,
it must have an unusually grey extinction curve. Otherwise, the
distribution must be patchy. However, these analyses also 
raise questions of how to account self-consistently for all the
observations. For example, the absorption inferred from 
the H$\alpha$/H$\beta$ ratio in starburst galaxies is typically
twice as high as that inferred from the ultraviolet spectral 
slope. One interpretation of this result is that the ionized gas
and ultraviolet-bright stars have different spatial distributions
(Calzetti 1997). Issues such as this highlight the need for a
simple yet versatile model to interpret simultaneously a wide 
range of phenomena related to the absorption of starlight by 
dust in galaxies. The purpose of this paper is to present such
a model.

We begin with the conventional view that young stars ionize H{\sc
ii} regions in the interiors of the dense clouds in which they 
are born. Line photons produced in the H{\sc ii} regions and the
non-ionizing continuum photons from young stars are absorbed in
the same way by dust in the outer H{\sc i} envelopes of the birth
clouds and the ambient ISM. The birth clouds, however, have 
finite lifetimes. Thus, non-ionizing ultraviolet and optical 
photons from stars that live longer than the birth clouds are 
absorbed only by the ambient ISM. This allows the 
ultraviolet continuum to be less attenuated than the emission 
lines. Our model builds on several previous studies. For example,
Silva et al.  (1998) considered the effects of finite lifetimes
of stellar birth clouds on the continuum but not the line emission
from galaxies. Here, we treat the transfer of radiation (especially 
scattering) in an approximate way, which would preclude a 
detailed description of surface brightnesses but should be 
appropriate for angle-averaged quantities such as luminosities.
Our model succeeds in accounting quantitatively for all the 
available observations of a homogeneous sample of nearby starburst 
galaxies. 

We present our model in \S2, where we express the effective
absorption curve describing the global transmission of radiation
in terms of the different components of the ISM. In \S3, we 
compare our model with observations and identify the specific
influence of each parameter on the different integrated spectral
properties of galaxies. One outcome of our detailed analysis
is a remarkably simple recipe for absorption, which provides a
good approximation to the observed mean relations and is easy
to incorporate into any population synthesis model. In \S4, we
explore how the spatial distribution of dust can be constrained
by the observations. Our conclusions are summarized
in \S5.

\section{Definition of the Model}

We first introduce some notation and nomenclature that will help
specify our model. The luminosity per unit wavelength
$L_\lambda(t)$ emerging at the time $t$ from a galaxy illuminated
by an internal stellar population can be expressed generally as
\begin{equation}
L_\lambda(t)=\int_0^t\,dt'\,\psi(t-t')\,S_\lambda(t')\,T_\lambda(t,t')\,.
\label{basic}
\end{equation}
Here $\psi(t-t')$ is the star formation rate at the time $t-t'$, 
$S_\lambda(t')$ is the luminosity emitted per unit wavelength and
per unit mass by a stellar generation of age $t'$, and $T_\lambda(t,t')$
is the ``transmission function,'' defined as the fraction of the radiation
produced at the wavelength $\lambda$ at the time $t$ by a generation of stars
of age $t'$ that escapes from the galaxy. Thus, $T_\lambda(t,t')$ must be
regarded as the average transmission along rays emanating in all directions
from all stars of age $t'$ within the galaxy. Since absorption and 
scattering by dust will cause the radiation to emerge anisotropically to
some degree, $L_\lambda$ can only be measured by a hypothetical set of
detectors completely surrounding the galaxy. In practice, this means that 
equation~(\ref{basic}) gives the mean of $4\pi D^2 f_\lambda$, where
$D$ is the distance and $f_\lambda$ is the observed flux, for an ensemble
of randomly oriented but otherwise similar galaxies.

For some purposes, it is convenient to reexpress $L_\lambda(t)$ as the
product of a mean transmission function weighted by the luminosity
of each stellar generation $\overline{T}_\lambda(t)$ and the
total unattenuated stellar luminosity:
\begin{equation}
L_\lambda(t)=\overline{T}_\lambda(t)\,\int_0^t\,dt'\,\psi(t-t')\,S_\lambda(t')\,,
\label{basic2}
\end{equation}
\begin{equation}
\overline{T}_\lambda(t)\equiv{
{\int_0^t\,dt'\,\psi(t-t')\,
S_\lambda(t')\,T_\lambda(t,t')}\over
{ \int_0^t\,dt'\,\psi(t-t')\,S_\lambda(t')}}\,.
\label{ttrans}
\end{equation}
We often reexpress $\overline{T}_\lambda(t)$ in terms of an ``effective
absorption'' optical depth $\hat{\tau}_\lambda^{\rm }(t)$ indicated
by a caret and defined by
\begin{equation}
\overline{T}_\lambda(t)\equiv \exp\left[-\hat{\tau}_\lambda^{\rm }(t)\right]\,.
\label{taueff}
\end{equation}
This nomenclature emphasizes that the luminosity of a galaxy is diminished
purely by the absorption of photons.\footnote{The optical depth 
$\hat{\tau}_\lambda^{\rm }$ defined by equation~(\ref{taueff}) has also
been referred to by several other names in the literature, including the optical
depth of ``apparent extinction,'' ``effective extinction,'' ``attenuation,''
``obscuration,'' and ``absorption.''} However, since scattering may 
also affect the path lengths of the photons through the ISM before they are
absorbed, $\hat{\tau}_\lambda^{\rm }(t)$ will differ in general from the
true absorption curve determined by the optical properties of the dust grains.
The fraction of stellar radiation not transmitted by the ISM, 
$1-\overline{T }_\lambda(t)$, is absorbed by dust. Using equation~(\ref{basic2})
and integrating over all wavelengths, we obtain the total luminosity absorbed
and reradiated by dust
\begin{equation}
L_{\rm dust}(t)=\int_0^\infty\,d\lambda\,\left[
1-\overline{T}_\lambda(t)\right]\,
\int_0^t\,dt'\,\psi(t-t')\,S_\lambda(t')\,.
\label{ldust}
\end{equation}

\begin{figure*}
\epsfxsize=8cm
\centerline{\epsfbox{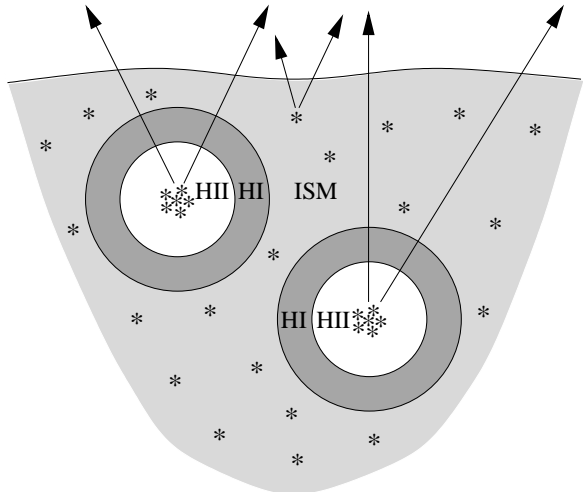}}
\figcaption{\small Schematic representation of the birth cloud and
ambient ISM surrounding each generation of stars in a model galaxy (see text).
Rays leaving in different directions are also shown.}
\end{figure*}

We now specify our model for computing the above properties of 
galaxies. This amounts to specifying the transmission function 
$T_\lambda(t,t')$. Our goal is to parameterize only those features
of the ISM that are essential to understanding the main physical
effects of dust on the integrated spectral properties of galaxies.
We adopt the simple but fairly conventional view illustrated in 
Figure~1. Young stars ionize H{\sc ii} regions in the inner parts
of the dense (mostly molecular) clouds in which they are born.
Line photons produced in the H{\sc ii} regions and the non-ionizing
continuum photons from young stars propagate through the outer H{\sc
i} envelopes of these ``birth clouds'' and then through the ``ambient
ISM'' before they escape from the galaxy. Both the H{\sc ii} and H{\sc
i} regions of the birth clouds may contain dust, and in the ambient
ISM, the dust may be distributed in a smooth or patchy way.
The birth clouds, however, have finite lifetimes. Thus, photons
emitted by longer-lived stars propagate only through the ambient 
ISM---where they are treated in the same way as photons emerging
from the birth clouds---before they escape from the galaxy. To
incorporate these basic features into our model, we make the
following simplifying assumptions. 

We assume that stars form in clusters with the galaxy-wide 
stellar initial mass function (IMF). This should be an excellent 
approximation for the purpose of computing the integrated spectral 
properties of a galaxy, although it could break down in clusters or 
regions with small numbers of stars. In our model, all star 
clusters are embedded in their birth clouds for some time and
then disrupt them or migrate away from them into the ambient ISM of
the galaxy. We assume for simplicity that the effective absorption 
in the birth clouds depends only on the stellar age $t'$ and is
constant in the ambient ISM. Hence, we can write the total 
transmission function in the form
\begin{equation}
T_\lambda(t,t')=T_\lambda^{\rm BC}(t')\,T_\lambda^{\rm ISM}\,,
\label{tsingle}
\end{equation}
where $T_\lambda^{\rm BC}$ and $T_\lambda^{\rm ISM}$ are the 
transmission functions of the birth clouds and the ambient ISM, 
respectively. In our model, the birth clouds, assumed to be 
identical, consist of an inner H{\sc ii} region ionized by young
stars and bounded by an outer H{\sc i} region. For simplicity, we 
assume that all stars remain within their birth clouds with a 
constant transmission function for a time $t_{\rm BC}$ and are 
then introduced immediately into the ambient ISM; thus
\begin{equation}
T_\lambda^{\rm BC}(t')=\cases
{T_\lambda^{\rm HII}\,T_\lambda^{\rm HI}\,,&for
$t'\leq t_{\rm BC}\,$,\cr
1\,,&for $t'>t_{\rm BC}\,$,\cr}
\label{tbc}
\end{equation}
where $T_\lambda^{\rm HII}$ and $T_\lambda^{\rm HI}$ are the 
transmission functions of the H{\sc ii} and H{\sc i} regions,
respectively. By writing the transmission functions as products
in equations (\ref{tsingle}) and (\ref{tbc}), we have assumed that
the propagation of photons in the H{\sc ii} and H{\sc i} regions
and the ambient ISM are independent of each other. This
prescription is exact for forward scattering and should be
a good approximation for the purpose of computing luminosities
even with some backward scattering.

An alternative but equivalent way of expressing the transmission
functions of the H{\sc ii} and H{\sc i} regions and the ambient 
ISM is in terms of the corresponding effective absorption optical depths,
$\hat{\tau}_\lambda^{\rm HII}$, $\hat{\tau}_\lambda^{\rm HI}$, and
$\hat{\tau}_\lambda^{\rm ISM}$. For non-ionizing photons (with
$\lambda>\lambda_L=912\,${\AA}), we thus write
\begin{equation}
T_\lambda^{\rm HII}=\exp(-\hat{\tau}_\lambda^{\rm HII})\,,
\label{thii}
\end{equation}
\begin{equation}
T_\lambda^{\rm HI}=\exp(-\hat{\tau}_\lambda^{\rm HI})\,,
\label{thi}
\end{equation}
\begin{equation}
T_\lambda^{\rm ISM}=\exp(-\hat{\tau}_\lambda^{\rm ISM})\,.
\label{tism}
\end{equation}
For some purposes, it is also convenient to define the total effective
absorption optical depth of dust in the birth clouds and the fraction of this
contributed by the internal H{\sc ii} regions
\begin{equation}
\hat{\tau}_\lambda^{\rm BC}=\hat{\tau}_\lambda^{\rm HII}+\hat{\tau}_\lambda^{\rm HI}\,,
\label{taubc}
\end{equation}
\begin{equation}
f=\hat{\tau}_\lambda^{\rm HII}/\left(\hat{\tau}_\lambda^{\rm HII}+
\hat{\tau}_\lambda^{\rm HI}\right)\,.
\label{fhii}
\end{equation}
We note that $f$ may or may not coincide with the ratio of the H{\sc ii}
to total gas column densities of the birth clouds, depending on whether
or not the dust-to-gas ratios in the H{\sc ii} regions are the same as
those in the H{\sc i} regions.

Finally, we assume that the emission lines are produced only in the
H{\sc ii} regions of the birth clouds. The justification for this is
that the lifetimes of the birth clouds are generally greater than
the lifetimes of the stars producing most of the ionizing photons, i.e.
about $3\times10^6\,$yr. In this case, line photons are absorbed in
the H{\sc i} regions and the ambient ISM in the same way as non-ionizing
continuum photons. The emergent luminosity of a line of wavelength
$\lambda_l$ is thus
\begin{equation}
L_l(t)=\int_0^{t_{\rm BC}}\,dt'\,\psi(t-t')\,S_l(t')
T_{\lambda_l}^{\rm HI}T_{\lambda_l}^{\rm ISM}\,,
\label{emline}
\end{equation}
where $S_l(t')$ is the luminosity of the line produced in the H{\sc
ii} regions by a stellar generation of age $t'$, and $T_{\lambda_l}^{\rm HI}$ 
and $T_{\lambda_l}^{\rm ISM}$ are given by equations (\ref{thi}) and (\ref{tism}).
The luminosities of H-recombination lines are proportional to the ionization 
rate,
\begin{equation}
\dot{N}_{\rm ion}= {{T_{\rm ion}}\over{hc}}\,
\int_0^{\lambda_L} d\lambda\,\lambda\,S_\lambda\,,
\label{nlyc}
\end{equation}
where $T_{\rm ion}$ is the fraction of Lyman continuum photons
absorbed by the gas rather than the dust in the H{\sc ii} regions. 
We adopt the formula
\begin{equation}
T_{\rm ion}=
{
{(\hat{\tau}_{\lambda_L}^{\rm HII})^3\exp(-\hat{\tau}_{\lambda_L}^{\rm HII})}\over
{3\lbrace(\hat{\tau}_{\lambda_L}^{\rm HII})^2 -2\hat{\tau}_{\lambda_L}^{\rm HII}+
2[1-\exp(-\hat{\tau}_{\lambda_L}^{\rm HII})]\rbrace}}\,.
\label{absion}
\end{equation}
This equation, which is equivalent to equation~(5-29) and Table~5.4 of 
Spitzer (1978), was derived by Petrosian, Silk, \& Field (1972) and shown by
them to provide a good approximation to the results of more detailed numerical 
computations.

We are especially interested in the emergent luminosities of the
H$\alpha$ and H$\beta$ Balmer lines of hydrogen, $L_{{\rm H}\alpha}$ and
$L_{{\rm H}\beta}$. We assume that, given $\dot{N}_{\rm ion}$, the 
production rates of all H-recombination photons are the same as for the
dust-free case~B recombination. This is validated by detailed photoionization
models of H{\sc ii} regions, including the effects of internal dust (Hummer \& 
Storey 1992; Ferland 1996; Bottorff et al. 1998). For electronic densities
$n_e\ll 10^4\,$cm$^{-3}$ and temperatures $T_e\approx10^4\,$K, we then have
(Osterbrock 1989)
\begin{eqnarray}
&S_{{\rm H}\alpha}=&0.450\,(hc/\lambda_{{\rm H}\alpha})\,\dot{N}_{\rm ion}\,,\cr
&S_{{\rm H}\beta}=&0.117\,(hc/\lambda_{{\rm H}\beta})\,\dot{N}_{\rm ion}\,,
\label{hlums}
\end{eqnarray}
with $\lambda_{{\rm H}\alpha}=6563\,${\AA} and $\lambda_{{\rm H}\beta}=4861
\,${\AA}. It is straightforward to extend this procedure to compute the 
luminosity of any other H-recombination line (e.g., P$\beta$, B$\gamma$).
We assume that all the power radiated in the form of L$\alpha$ photons,
$\tilde{S}_{{\rm L}\alpha}=0.676\,(hc/\lambda_{{\rm L}\alpha})\,\dot{N}_{\rm
ion}$, is eventually absorbed by dust as a consequence of resonant 
scattering ($\lambda_{L \alpha}=1216\,${\AA}). We have checked that our
conclusions are not affected even if a significant fraction of L$\alpha$ 
photons escapes from the galaxy. For completeness, we compute the power from
ionizing radiation that is neither absorbed by dust nor emerges in the H$\alpha$
and H$\beta$ lines,
\begin{equation}
S_X=
\bigg(T_{\rm ion}\,\int_0^{\lambda_L} d\lambda\,S_\lambda\bigg)-
\left(\tilde{S}_{{\rm L}\alpha}+S_{{\rm H}\alpha}+S_{{\rm H}\beta}\right)\,,
\label{lx}
\end{equation}
and distribute this uniformly in wavelength between 3000 and 6000~{\AA} (a
range that includes most of the relevant emission lines). This is only a few 
percent of all the stellar radiation and has a negligible influence on our
results.

We now consider how the transmission function of the ambient ISM depends
on the spatial distribution and optical properties of the dust. A 
particularly enlightening way to express $T_\lambda^{\rm ISM}$ is in terms
of the probability density for the absorption of photons emitted in all
directions by all stars within a galaxy. We define $\tau_\lambda$
as the optical depth of absorption along the path of a photon, from its
emission by a star to its escape from the galaxy, including the influence
of scattering on the path. From now on, we refer to $\tau_\lambda$
simply as the ``optical depth'' of the dust.\footnote{The optical depth
denoted here by $\tau_\lambda$ is sometimes referred to as an effective 
optical depth and denoted by $\tau_*$ (Rybicki \& Lightman 1979). In this
paper, we use the term ``effective optical depth'' to refer to the
geometry-dependent optical depth $\hat{\tau}_\lambda$ defined by
equation~(\ref{taueff}).} We further define $p(\tau_\lambda) 
d\tau_\lambda$ as the probability that the optical depth lies between 
$\tau_\lambda$ and $\tau_\lambda + d\tau_\lambda$. Integrating over rays
emanating in all directions from all stars within the galaxy, we can 
then write the transmission function of the ambient ISM as
\begin{equation}
T_\lambda^{\rm ISM}=\int_0^\infty\,d\tau_\lambda\,p(\tau_\lambda)\exp(-\tau_\lambda)\,.
\label{tlamism}
\end{equation}
If the spatial distribution of the dust is specified, this determines 
$p(\tau_\lambda)$, and $T_\lambda^{\rm ISM}$ then follows from
equation~(\ref{tlamism}). For some purposes, however, it is more 
informative to specify $p(\tau_\lambda)$ itself rather than any of
the spatial distributions of the dust that could give rise
to it when evaluating $T_\lambda^{\rm ISM}$.

The optical depth $\tau_\lambda$ of the dust depends on a combination
of absorption and scattering. If scattering were only in the forward 
direction, $\tau_\lambda$ would equal the true absorption optical depth.
In the case of isotropic scattering, $\tau_\lambda$ is given by the 
geometrical mean of the true absorption and extinction (absorption plus
scattering) optical depths, i.e.,
\begin{equation}
\tau_\lambda=\left[\tau_\lambda^a(\tau_\lambda^a+\tau_\lambda^s)\right]^{1/2}
=\tau_\lambda^a (1- \omega_\lambda)^{-1/2}\,,
\label{tscat}
\end{equation}
where $\omega_\lambda$ is the single-scattering albedo (Rybicki \& 
Lightman 1979).\footnote{Equation~(\ref{tscat}) is exact in the limit
of large extinction optical depths but overestimates $\tau_\lambda$ 
in the opposite limit.} Although this formula was derived 
for an infinite homogeneous medium, it also provides an excellent 
approximation for more complicated geometries (see, e.g., Fig.~5
of Silva et al. 1998, which compares this approximation with a Monte
Carlo calculation by Witt, Thronson, \& Capuano 1992). In reality, 
both observations and models suggest that interstellar dust in the 
Milky Way has pronounced forward scattering at ultraviolet and optical 
wavelengths (e.g., Draine \& Lee 1984; Kim, Martin, \& Hendry 1994; 
Calzetti et al. 1995). Thus, we expect the truth to lie between the 
idealized formulae $\tau_\lambda= \tau_\lambda^a$ and $\tau_\lambda=
\tau_\lambda^a(1- \omega_\lambda)^{ -1/2}$. In our model, the 
details of scattering and absorption are subsumed in our approximation
of the optical depth of the dust as a power law in wavelength, 
$\tau_\lambda \propto \lambda^{-m}$ (\S4). This approach would 
be questionable if we wished to compute directional quantities
such as surface brightnesses. However, since we compute only
angle-averaged quantities (i.e., luminosities), our simple treatment
of scattering should be an adequate approximation. The reason
for this is that, since many of
the photons scattered out of a line of sight will be replaced by
other photons scattered into it, inaccuracies in the treatment of 
scattering will partially (but not totally) cancel out when the 
emission is averaged over all directions.

We now consider three specific, idealized distributions of
dust in the ambient ISM: a uniform foreground screen, a mixed slab,
and discrete clouds. These are fairly standard representations, which 
may be regarded as illustrative of a wide range of possibilities. Other 
models could be constructed for more realistic and hence more
complicated distributions of dust, for example by multiplying
the transmission functions derived from different $p(\tau_\lambda)$'s.
It should be recalled in any case that the total transmission 
function in our model is the product of that for the birth clouds
and that for the ambient ISM, as specified by equation~(\ref{tsingle}).

{\it Foreground screen}$\,$. The simplest model is that of a 
uniform foreground screen, which absorbs the light from all stars
by the same amount. While unrealistic, this model has been adopted
in many previous studies, and we include it here for reference. We 
call $\tau_\lambda^{\rm sc}$ the optical depth of the screen. We then
have
\begin{equation}
p(\tau_\lambda)=\delta(\tau_\lambda-\tau_\lambda^{\rm sc})\,,
\label{pscreen}
\end{equation}
\begin{equation}
T_\lambda^{\rm ISM}=\exp(-\tau_\lambda^{\rm sc})\,,
\label{tscreen}
\end{equation}
where $\delta$ is the Dirac delta function.

{\it Mixed slab}$\,$. A potentially more realistic model is 
that of a slab filled with a uniform mixture of sources and absorbers
(i.e., stars and dust). This can account for differences in the 
absorption affecting stars near the midplane or near the surface of 
a galactic disk. We call $\tau_\lambda^{\rm sl}$ the full optical 
depth through the slab in the direction normal to the surface. By 
considering sources located at all points within the slab and rays 
emanating at all angles from the normal, we obtain, after some 
manipulation,
\begin{equation}
p(\tau_\lambda)=\cases
{1/(2\tau_\lambda^{\rm sl})\,,&for
$\tau_\lambda\leq \tau_\lambda^{\rm sl}\,$,\cr
\tau_\lambda^{\rm sl}/(2\tau_\lambda^2)\,,&for
$\tau_\lambda>\tau_\lambda^{\rm sl}\,$,\cr}
\label{pslab}
\end{equation}
\begin{equation}
T_\lambda^{\rm ISM}={1\over{2\tau_\lambda^{\rm sl}}}
\left[1+(\tau_\lambda^{\rm sl}-1)\exp({-\tau_\lambda^{\rm sl}})
-(\tau_\lambda^{\rm sl})^2E_1(\tau_\lambda^{\rm sl})\right]\,,
\label{tslab}
\end{equation}
where $E_1$ is the exponential integral of the first order.

{\it Discrete clouds}$\,$. Both the foreground screen and mixed slab
models are based on smooth representations of the ambient ISM. In real galaxies,
however, the ISM is organized into clouds and other structures. The photons 
emitted in different directions by different stars can encounter any number
of these clouds on their way out of the galaxy. Thus, $p(\tau_\lambda)$ may
be better approximated by a Poisson distribution (Spitzer 1978; 
Natta \& Panagia 1984). We call $\tau_\lambda^{\rm c}$ the optical
depth per cloud, and $\overline{n}$ the mean number of clouds encountered
along different lines of sight.  The mean optical depth is thus 
$\overline{\tau}_\lambda= \overline{n} \tau_\lambda^{\rm c}$. In this 
case, we have\footnote{Strictly, equation~(\ref{ppois}) corresponds to a
Poisson distribution of identical, face-on screens. For spherical
clouds, the factor $\delta(\tau_\lambda-n \tau_\lambda^{\rm c})$ would have
to be replaced by a function with a finite width to account for the different
optical depths seen by photons with different impact parameters. We ignore
this subtlety here.}
\begin{equation}
p(\tau_\lambda)=\sum_{n=0}^\infty {{\,{\overline{n}}^n}\over{n!}}
                \exp(-\overline{n})
                \delta(\tau_\lambda-n\tau_\lambda^{\rm c})\,,
\label{ppois}
\end{equation}
\begin{equation}
T_\lambda^{\rm ISM}=\exp\left[-\overline{n}(1-e^{-\tau_\lambda^{\rm c}})\right]\,.
\label{tpois}
\end{equation}
In the limits $\tau_\lambda^{\rm c}\rightarrow0$ and $\tau_\lambda^{\rm c}
\rightarrow\infty$, we obtain the familiar expressions $T_\lambda^{\rm ISM}
\rightarrow\exp(-\overline{\tau}_\lambda)$ and $T_\lambda^{\rm ISM}
\rightarrow\exp(-\overline{n})$. This model should provide an adequate
description of the absorption along individual random lines of sight, as
required by equation~(\ref{tlamism}), but it neglects possible correlations 
between the absorption along multiple nearby lines of sight, which may be
important in other applications.

In principle, we could compute models with arbitrary star formation
rates. For simplicity, in all the following applications, we take the 
star formation rate to be constant and include the age $t$ of the stellar
population as a parameter. This should be regarded as the effective age
of the most recent burst of star formation, which can range from a few 
times $10^7\,$yr for a galaxy in an active starburst to several times 
$10^9\,$yr for a more quiescent galaxy. With the assumption $\psi={\rm
const}$, equations~(\ref{ttrans}), (\ref{taueff}), and 
(\ref{tsingle})--(\ref{taubc}) allow us to reexpress the effective 
absorption optical depth at the time $t$ in terms of the effective 
absorption optical depths in the birth clouds and the ISM as
\begin{equation}
\hat{\tau}_\lambda^{\rm }(t)=\hat{\tau}_\lambda^{\rm BC}+\hat{\tau}_\lambda^{\rm ISM}
			 +\Delta\hat{\tau}_\lambda^{\rm }(t)\,,
\label{newtaueff}
\end{equation}
with
\begin{eqnarray}
&&\Delta\hat{\tau}_\lambda^{\rm }(t)=\cr
&&\cr
&&\cases
{0,&for $t\leq t_{\rm BC}\,$,\cr
-\ln\left[{{\int_0^{t_{\rm BC}}dt'\,S_\lambda(t')\,+\,
              e^{\hat{\tau}_\lambda^{\rm BC}}\int_{t_{\rm BC}}^t dt'\,S_\lambda(t')}\over
             {\int_0^t dt'\,S_\lambda(t')}}\right],
&for $t> t_{\rm BC}$.\cr}\cr
&&
\label{deltataueff}
\end{eqnarray}
We compute $S_\lambda(t')$ using the latest version of the Bruzual \&
Charlot (1993) population synthesis code. Our conclusions would not
be affected by the use of other standard codes. Unless otherwise 
indicated, we adopt solar metallicity and a Salpeter IMF truncated at
0.1 and 100$\,M_\odot$.

We now have a completely specified model for the absorption of 
starlight by dust in galaxies. The main adjustable parameters
describing the birth clouds are: the lifetime of the clouds, $t_{\rm
BC}$; the wavelength dependence and normalization of $\hat{
\tau}_\lambda^{\rm BC}$; and the fraction of $\hat{\tau}_\lambda^{\rm
BC}$ contributed by dust in H{\sc ii} regions, $f$. For the ambient ISM, 
the parameters depend on whether we adopt a foreground screen model
($\tau_\lambda^{\rm sc}$), a mixed slab model ($\tau_\lambda^{\rm sl}$),
or a discrete cloud model ($\tau_\lambda^{\rm c}$ and $\overline{n}$).
The other main parameter is the effective age of the starburst,
$t$. We compute the continuum luminosity per unit wavelength emerging
from a galaxy using equations~(\ref{basic}) and 
(\ref{tsingle})--(\ref{tism}). The emergent H$\alpha$ and H$\beta$ line
luminosities, $L_{{\rm H}\alpha} (t)$ and $L_{{\rm H} \beta}(t)$, are
obtained from equations~(\ref{thi}), (\ref{tism}), and 
(\ref{emline})--(\ref{hlums}).  The ratio of the H$\alpha$ line and 
continuum luminosities is then the emission equivalent width of the
emergent H$\alpha$ line, $W_{{\rm H} \alpha} (t)$. Finally, we compute
the far-infrared luminosity of the galaxy, $L_{\rm dust}(t)$, using 
equations~(\ref{ttrans}) and (\ref{ldust}).

In the remainder of this paper, we take two approaches to 
compare our model with observations. As equations~(\ref{tism})
and (\ref{tlamism}) indicate, the optical properties and spatial 
distribution of the dust affect observable quantities only through
the effective absorption curve $\hat{\tau}_\lambda^{\rm ISM}$. 
Different combinations of $\tau_\lambda$ and $p(\tau_\lambda)$, 
however, could lead to the same wavelength dependence of $\hat{\tau
}_\lambda^{\rm ISM}$. Thus, it is of primary interest to study how 
the wavelength dependence of $\hat{\tau }_\lambda^{\rm ISM}$ is 
constrained by observations. This is our goal in \S3 below. 
Then, in \S4, we explore whether there are examples of optical 
properties and spatial distributions of the dust that satisfy these
constraints, recognizing that there is no unique solution.

\section{Comparisons with Observations: the Effective Absorption Curve}

In this section, we use observations of nearby starburst galaxies to 
constrain the wavelength dependence of the effective absorption curve
$\hat{\tau}_\lambda^{\rm }$ defined by equation~(\ref{taueff})
without specifying the spatial distribution and optical properties of
the dust. We also identify the specific influence of each parameter 
in the model on different integrated spectral properties of galaxies 
and investigate the physical origin of the observed relationships 
between these properties. We assume for simplicity that the effective
absorption curves of the birth clouds and the ambient ISM have the 
same power-law form,
\begin{equation}
\hat{\tau}_\lambda^{\rm BC}=\hat{\tau}_V^{\rm BC}
( \lambda / 5500\,\hbox{\AA})^{-n}\,,
\label{lambc}
\end{equation}
\begin{equation}
\hat{\tau}_\lambda^{\rm ISM}=\hat{\tau}_V^{\rm ISM}
(\lambda/5500\,\hbox{\AA})^{-n}\,,
\label{lamism}
\end{equation}
where $n$ is a parameter to be determined, and $\hat{\tau}_V^{\rm BC}$
and $\hat{\tau}_V^{\rm ISM}$ are the normalizing coefficients at 
5500~{\AA}. 

To calibrate our model, we appeal to a homogeneous set of observations
of nearby ultraviolet-selected starburst galaxies, for
which the ultraviolet, far-infrared,
H$\alpha$, and H$\beta$ fluxes, H$\alpha$ equivalent widths,
and ultraviolet spectral slopes are available. The sample, compiled
by Meurer et al. (1999), includes 57 non-Seyfert galaxies from the
{\it International Ultraviolet Explorer} ({\it IUE}) atlas of Kinney
et al. (1993) classified into one of the starburst categories (i.e.,
starburst nucleus, starburst ring, blue compact dwarf galaxy, or blue
compact galaxy). This sample has been further restricted by Meurer et 
al. (1999) to galaxies with optical angular diameters less than 4', for
which most of the (usually concentrated) ultraviolet emission could be
observed within the {\it IUE} aperture. The galaxies span a wide range
of morphological types, from Sab to Irr, and a wide range of 
absolute $B$-band magnitudes, $-16\la M_B\la -22$ (we adopt a Hubble 
constant $H_0= 70 $~km$\,$s$^{-1}$Mpc$^{-1}$). From the {\it IUE} 
spectra of all galaxies in the sample, Meurer et al. (1999) compute 
the restframe ultraviolet flux at 1600~{\AA}, $F_{1600}$, and the slope
$\beta$ of the ultraviolet continuum. The quantity $F_{1600}$
is defined as $\lambda f_\lambda$, where $f_{1600}$ is the mean flux
density over a square passband with a central wavelength of 1600~{\AA}
and a width of 350~{\AA}. The slope $\beta$ is determined from a 
power-law fit of the form $f_\lambda \propto \lambda^\beta$ to the 
observed ultraviolet spectrum in ten continuum bands that avoid strong
stellar and interstellar absorption features (in particular, the 
2175~{\AA} dust feature) in the range $1268\leq\lambda\leq
2580\,${\rm {\AA}}, as defined by Calzetti et al. (1994). These 
ultraviolet properties were derived for galaxies in the sample after
removing foreground extinction by the Milky Way.

Observations with the {\it Infrared Astronomical Satellite} 
({\it IRAS}) are available for 47 galaxies in the sample. To estimate
the total far-infrared flux $F_{\rm dust}$ from the observed {\it IRAS} flux 
densities, $f_{\nu}(60\,\micron)$ and $f_{\nu}(100\,\micron)$, we follow
the prescription of Meurer et al. (1999). They first calculate the 
quantity $F_{\rm FIR}=1.26\times 10^{-11} [2.58f_{\nu}(60\,\micron)+
f_{\nu} (100\,\micron)]\,\,$ergs$\,$cm$^{-2}$s$^{-1}$ defined by Helou
et al. (1988) and then compute $F_{\rm dust}$ ($F_{\rm bol}$ in their 
notation) from $F_{\rm FIR}$ and $f_{\nu}(60 \,\micron) /f_{\nu} 
(100\,\micron)$ using the bolometric correction 
appropriate for dust with a single 
temperature and an emissivity proportional to frequency $\nu$. This 
prescription agrees reasonably well with observations from 
near-infrared to submillimeter wavelengths for a small sample of 
galaxies (see Fig.~2 of Meurer et al. 1999). The corrections from 
$F_{\rm FIR}$ to $F_{\rm dust}$ are relatively small, with a mean 
$\langle F_{\rm dust} /F_{\rm FIR}\rangle \approx1.4$ and a standard 
deviation of 0.1. Finally, for 31 galaxies, the ratio of H$\alpha$
to H$\beta$ fluxes and the 
H$\alpha$ equivalent width are available from 
Storchi-Bergmann, Kinney, \& Challis (1995) and McQuade, Calzetti, \& 
Kinney (1995), with corrections for stellar absorption by Calzetti
et al. (1994).  Neglecting possible anisotropies, we equate 
flux ratios at all wavelengths to the corresponding luminosity ratios. 
Figures~2, 3, and 4 show $L_{\rm dust}/ L_{1600}$, $L_{{\rm H} 
\alpha}/L_{{\rm H}\beta}$, and $W_{{\rm H} \alpha}$, respectively, as
a function of $\beta$ for this sample. The typical scatter at fixed 
$\beta$ amounts to a factor of about 2 in $L_{\rm dust} /L_{1600}$ and
$L_{{\rm H}\alpha}/ L_{{\rm H}\beta}$, but nearly a factor of 10 in 
$W_{{\rm H}\alpha}$. Thus, both $L_{\rm dust}/ L_{1600}$ and 
$L_{{\rm H}\alpha}/ L_{{\rm H} \beta}$, to a first approximation, 
are functions of $\beta$, while there is little or no correlation between
$W_{{\rm H}\alpha}$ and $\beta$. We note, however, that the scatter 
even in the first two relations appears to be real because it is 
larger than the typical measurement errors (which are shown in the 
upper left panels of Figs.~2--4).

\begin{figure*}
\epsfxsize=12cm
\centerline{\epsfbox{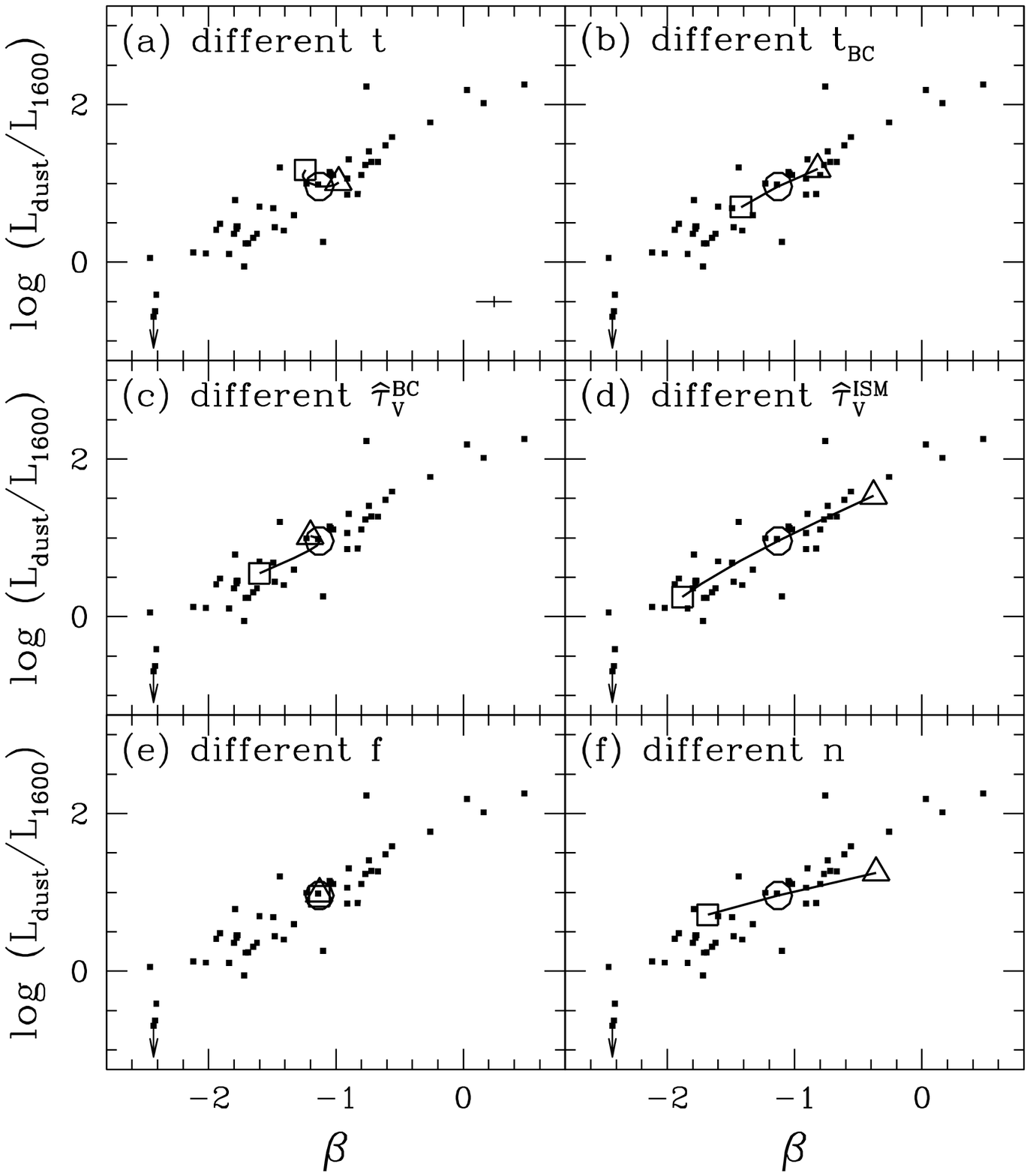}}
\figcaption[f2.eps]{\small Ratio of far-infrared to ultraviolet luminosities plotted
against ultraviolet spectral slope. The data points are from the Meurer et
al. (1999) sample discussed in \S3 and are repeated in all panels (typical 
measurement errors are indicated in the upper left panel). In each case, the 
line shows the effect of varying one parameter from the lower end of the range
({\it square}) to the standard value ({\it circle}) to the upper end of the
range ({\it triangle}), with all other parameters fixed at their standard values
(eq.~[\ref{standard}]): ({\it a}) effective starburst ages, $t=3\times10^7$, 
$3\times10^8$, and $3\times10^9\,$yr; ({\it b}) lifetimes of the birth clouds,
$t_{\rm BC}=3\times10^6$, $1\times10^7$, and $3\times10^7\,$yr; ({\it c}) 
effective optical depths in the birth clouds, $\hat{\tau}_V^{\rm BC}=0.0$, 1.0, and 
2.0; ({\it d}) effective optical depths in the ambient ISM, $\hat{\tau}_V^{\rm 
ISM}=0.0$, 0.5, and 1.0; ({\it e}) fractions of dust in H{\sc ii} regions in the
birth clouds, $f=0.0$, 0.1, and 1.0; and ({\it f}) exponents in the relation 
$\hat{\tau}_\lambda^{\rm ISM}\propto\lambda^{-n}$, $n=0.4$, 0.7, and 1.0.}
\end{figure*}

\begin{figure*}
\epsfxsize=12cm
\centerline{\epsfbox{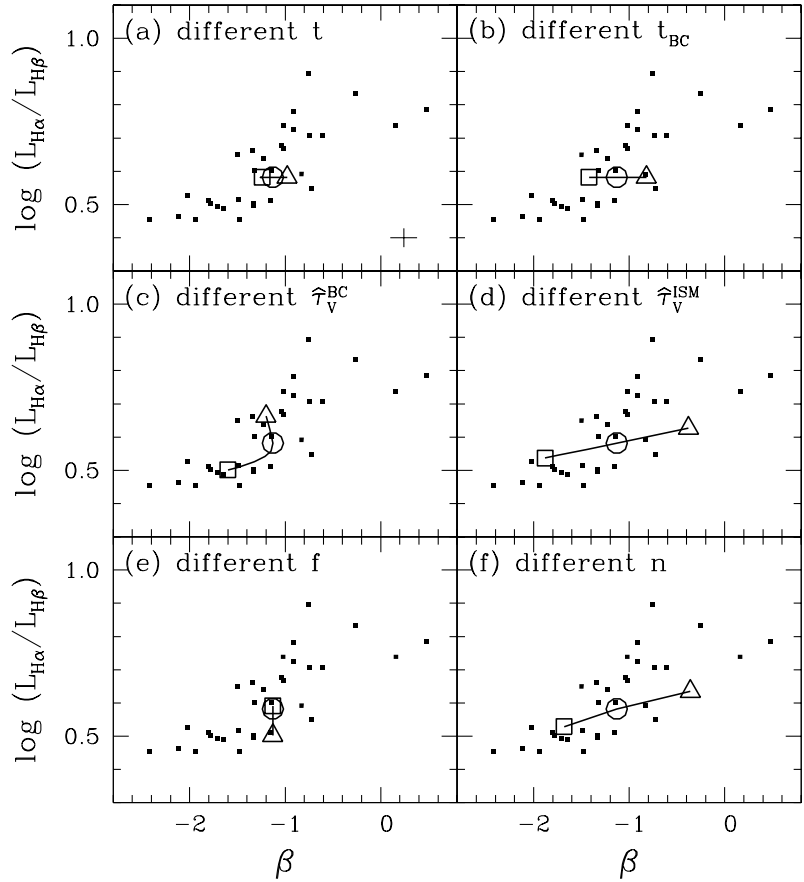}}
\figcaption[f3.eps]{\small H$\alpha/$H$\beta$ ratio plotted against ultraviolet spectral
slope. The data points are from the Meurer et al. (1999) sample discussed in
\S3 and are repeated in all panels (typical measurement errors are indicated in
the upper left panel). In each case, the line shows the effect of varying 
one parameter from the lower end of the range ({\it square}) to the standard
value ({\it circle}) to the upper end of the range ({\it triangle}), 
with all other parameters fixed at their standard values (eq.~[\ref{standard}]).
The models are the same as in Fig.~2.}
\end{figure*}
 
\begin{figure*}
\epsfxsize=12cm
\centerline{\epsfbox{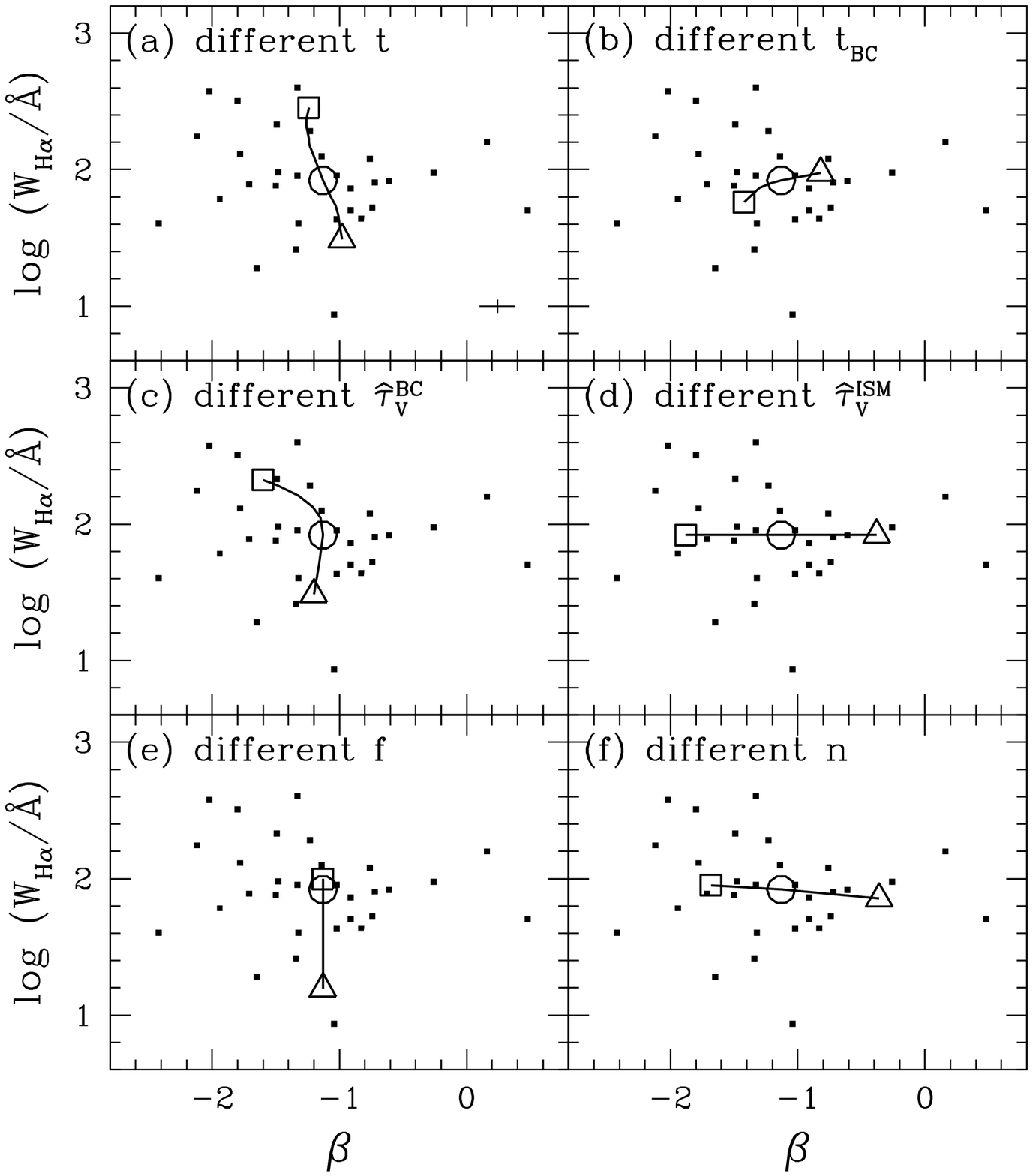}}
\figcaption[f4.eps]{\small H$\alpha$ equivalent width plotted against ultraviolet
spectral slope.  The data points are from the Meurer et al. (1999) sample 
discussed in \S3, with H$\alpha$ equivalent widths from Storchi-Bergmann et al.
(1995) and McQuade et al. (1995), and are repeated in all panels (typical 
measurement errors are indicated in the upper left panel). In each case, the 
line shows the effect of varying one parameter from the lower end of the range
({\it square}) to the standard value ({\it circle}) to the upper end
of the range ({\it triangle}), with all other parameters fixed at their 
standard values (eq.~[\ref{standard}]). The models are the same as in Fig.~2.}
\end{figure*}

We now use these observations to constrain the parameters in our model.
Our goal is to identify models that account for the typical properties,
scatter, and trends seen in the sample, rather than to interpret
in detail the emission from individual galaxies. Such an attempt might
be of limited validity because the predictions of our model, unlike the
observations, are averaged over angles.\footnote{Although some differences
are expected between the luminosity ratios and flux ratios due to 
anisotropic emission from galaxies, these must be relatively small or
else the relations shown in Figures~2 and 3 would be washed out. The
obvious outlier at $\beta =-0.76$ and $\log(L_{\rm dust}/ L_{1600})=2.2$ in
Figure~2 could be a galaxy for which some of the ultraviolet emission was
missed with {\it IUE} (see the discussion in \S3.1 in Meurer et al. 1999).}
We have checked, however, that the properties of every galaxy in the sample
could be reproduced with at least one combination of parameters. The 
influence of each parameter on observable quantities can best be
explored by keeping all other parameters fixed at ``standard'' values.
After some experimentation, we adopted:
\begin{eqnarray}
t&=&3\times10^8\,{\rm yr}\cr
t_{\rm BC}&=&1\times10^7\,{\rm yr}\cr
\hat{\tau}_V^{\rm BC}&=&1.0\cr
\hat{\tau}_V^{\rm ISM}&=&0.5\cr
f&=&0.1\cr
n&=&0.7\,.
\label{standard}
\end{eqnarray}
While these values are not the results of a rigorous optimization 
procedure, they do enable the standard model to match roughly the 
observed typical (i.e., median) properties of the starburst sample.
We note that $\hat{\tau}_V^{\rm BC}\approx1$ is typical of the 
observed optical depths to O-type and supergiant stars in the Milky
Way (Humphreys 1978). The dense cores of molecular clouds
have higher optical depths, but even short-lived stars spend only the
first 10\%$-$20\% of their lifetimes in these regions (e.g., Mezger
\& Smith 1977). Also, $t_{\rm BC}\approx10^7\,$yr is typical of the
timescale for the dispersal of giant molecular clouds in the Milky
Way (Blitz \& Shu 1980).

We now compare the predictions of our model with the observations 
described above.  Each panel in Figures~2--4 shows the effect of 
increasing and decreasing one parameter with respect to its standard
value with the others held fixed. We can summarize the role of
each parameter as follows.

{\it Effective starburst age}$\,$. The effect of increasing $t$ is most
noticeable on the H$\alpha$ equivalent width, which drops from almost 
300~{\AA} at $t= 3\times 10^7 \,$yr to less than 100~{\AA} at $t= 3\times
10^8\,$yr because of the rising contribution by old stars to the continuum 
near the line (Fig.~4{\it a}). Increasing $t$ also makes the ultraviolet 
spectral slope larger because old stars have intrinsically red
spectra. However, this does not affect the H$\alpha/$H$\beta$ ratio, 
since the emission lines are produced mainly by stars younger than
about $3\times 10^6 \,$yr (Fig.~3{\it a}).

{\it Lifetime of the birth clouds}$\,$. Increasing $t_{\rm BC}$ makes both
$\beta$ and the ratio of far-infrared to ultraviolet luminosities larger 
because the radiation from young blue stars is heavily absorbed for a 
longer time (Fig.~2{\it b}). Changes in $t_{\rm BC}$ do not affect 
$L_{{\rm H}\alpha}/L_{{\rm H}\beta}$ because we have assumed that the birth
clouds last longer than $3\times 10^6 \,$yr, the lifetime of stars producing
most of the ionizing photons (\S2). However, since more of the 
continuum luminosity near H$\alpha$ is absorbed for larger $t_{\rm BC}$, 
the H$\alpha$ equivalent width also increases (Fig.~4{\it b}). 

{\it Effective absorption optical depth in the birth clouds}$\,$.
Increasing $\hat{\tau}_V^{\rm BC}$ above unity has the remarkable property of
making $L_{{\rm H}\alpha}/L_{{\rm H}\beta}$ larger at almost fixed $\beta$ 
and $L_{\rm dust}/L_{1600}$ (Figs.~2{\it c} and 3{\it c}). The reason for this
is that, while $L_{{\rm H}\alpha}/L_{{\rm H}\beta}$ traces the absorption
in the H{\sc i} envelopes of the birth clouds, the ultraviolet radiation 
from embedded stars is so attenuated for $\hat{\tau}_V^{\rm BC}\ga1$ that
$L_{\rm dust}$ saturates, and $\beta$ and $L_{1600}$ are dominated
by stars in the ambient ISM. This has fundamental observational implications
(see below). Figure~4{\it c} also shows that the H$\alpha$ equivalent width
is very sensitive to changes in $\hat{\tau}_V^{\rm BC}$, which affect $L_{{\rm
H} \alpha}$.

{\it Effective absorption optical depth in the ambient ISM}$\,$. Since 
$\hat{\tau}_V^{\rm ISM}$ controls the absorption of stars that have emerged
from their birth clouds, it has a major influence on the ultraviolet 
spectra and hence on $\beta$ and $L_{\rm dust}/L_{1600}$ 
(Fig.~2{\it d}). Changes in $\hat{\tau}_V^{\rm ISM}$ also affect
the H$\alpha/$H$\beta$ ratio, although with a strong dependence on $\beta$ 
(Fig.~3{\it d}). As expected, $\hat{\tau}_V^{\rm ISM}$ has no effect 
on $W_{{\rm H}\alpha}$.

{\it Fraction of dust in the HII regions}$\,$. Increasing $f$ causes more
ionizing photons to be absorbed by dust and hence reduces the H$\alpha$
equivalent width considerably (Fig.~4{\it e}). Dust in H{\sc ii} regions
has a negligible effect on the H$\alpha/$H$\beta$ ratio. However, since we
have kept $\hat{\tau}_V^{\rm BC}$ fixed, increasing $f$ reduces the 
absorption in the H{\sc i} envelopes of the birth clouds and hence the 
H$\alpha/$H$\beta$ ratio in Figure~3{\it e}. Changes in $f$ have no effect
on $\beta$ and $L_{1600}$ and a negligible effect on $L_{\rm dust}$.

{\it Exponent of the effective absorption curve}$\,$. Increasing $n$ steepens the
effective absorption curve. This increases the ratio of far-infrared to ultraviolet
luminosities, the H$\alpha/$H$\beta$ ratio, and the ultraviolet spectral slope and
reduces slightly the H$\alpha$ equivalent width (Figs.~2{\it f}--4{\it f}). We
elaborate below on the reason for adopting $n=0.7$ as a standard value.

{\it Other parameters}$\,$. We have also computed models with different
IMFs and metallicities (not shown). Changes in these parameters have
no influence on $L_{{\rm H}\alpha}/L_{{\rm H}\beta}$ and affect $\beta$ and
$L_{\rm dust}/ L_{1600}$ at the level of less than 20\%. The influence on
the H$\alpha$ equivalent width is slightly stronger. For example, adopting a
Scalo IMF instead of a Salpeter IMF would reduce $W_{{\rm H}\alpha}$ by
about 50\% for the standard model because of the smaller proportion of
massive stars. In contrast, adopting a metallicity 20\% of the solar
value would increase $W_{{\rm H}\alpha}$ by about 20\% because metal-poor
stars are hotter than solar-metallicity stars. We also tested the
influence of a possible leakage of some ionizing photons from H{\sc
ii} regions into the ambient ISM by allowing 50\% of the line photons 
produced by stars in the standard model to escape without absorption
in the H{\sc i} envelopes of the birth clouds (Oey \& Kennicutt 1997, 
and references therein). While the H$\alpha/$H$\beta$ ratio was found to
be only 12\% lower in this case than in the standard model, the H$\alpha$
equivalent width increased by 60\%.

\begin{figure*}
\epsfxsize=8cm
\centerline{\epsfbox{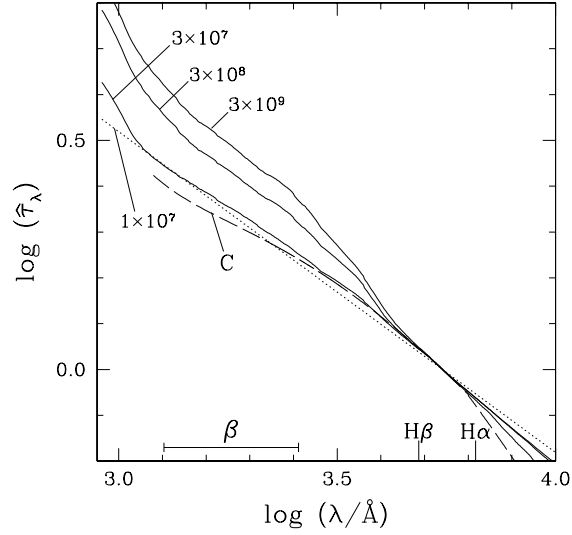}}
\figcaption[f5.eps]{\small Effective absorption curve (as defined by
eq.~[\ref{taueff}]) of our standard model at different effective starburst
ages, as indicated. The dashed curve is from Calzetti et al. (1994).  For
reference, the wavelength range where the ultraviolet slope $\beta$ is 
determined and the wavelengths of the H$\alpha$ and H$\beta$ lines are 
indicated at the bottom. All curves are normalized to unity at 5500~{\AA}.}
\end{figure*}
It is also interesting to examine the shape of the total effective 
absorption curve defined by equation~(\ref{taueff}) and its 
dependence on the age of the stellar population. In Figure~5,
we show $\hat{\tau}_\lambda^{\rm }(t)$ at different ages for our standard
model (computed from eqs.~[\ref{newtaueff}] and [\ref{deltataueff}]).
Initially, for $t\leq t_{\rm BC}$, $\hat{\tau }_\lambda^{\rm }$ is a 
power law of index $n=0.7$. At later times, the effective absorption
curve steepens markedly at wavelengths $\lambda \la 1200 \,${\AA}
($\log\lambda \la3.1$). It is more similar to the initial power
law in the wavelength range where $\beta$ is determined ($1200 \la
\lambda\la2500\,${\AA}), but steeper again at wavelengths $\lambda
\ga4000 \,${\AA} ($\log \lambda \ga3.6$). The reason for this is that
short-lived, massive stars, which produce most of the luminosity at
$\lambda \la 1200\,${\AA}, are embedded in the birth clouds, where their
radiation is heavily absorbed. Their contribution is substantial
and nearly constant across the wavelength range $1200\la \lambda\la2500\,
${\AA} but less important at longer wavelengths. The steepening of 
$\hat{\tau }_\lambda^{\rm }$ at $\lambda \ga4000 \,${\AA} is also 
enhanced by the appearance in the ambient ISM of long-lived stars with
a strong 4000~{\AA} break in their spectra, which contribute more light
redward than blueward of the break. The absorption in the birth 
clouds, therefore, has a major influence on the shape of the effective 
absorption curve.

It is worth pausing here to emphasize the important implications of
the finite lifetimes of the birth clouds. As mentioned in \S1, 
emission lines appear to be more attenuated than the continuum 
radiation in starburst galaxies. This led Calzetti (1997) to
suggest that ultraviolet-bright stars must be somehow physically
detached from the gas they ionize. In our model, the
lines produced in H{\sc ii} regions and the non-ionizing continuum
radiation from young stars are attenuated in the same way by dust
in the outer H{\sc i} envelopes of the birth clouds and the ambient
ISM. However, since the birth clouds have finite lifetimes, the 
non-ionizing continuum radiation from stars that live longer than
the birth clouds is attenuated only by the ambient ISM. The
ultraviolet and optical continuum radiation, therefore, appears 
to be less attenuated than the lines. This can be seen clearly in
Figures~3{\it c} and 4{\it c} from the rise in the H$\alpha /
$H$\beta$ ratio accompanying the drop in the H$\alpha$ equivalent
width at nearly constant ultraviolet spectral slope as the effective
absorption optical depth of the birth clouds increases above unity.
Thus, the finite lifetime of the birth clouds is a key 
ingredient to resolve the apparent discrepancy between the 
attenuation of line and continuum photons in starburst galaxies.

\begin{figure*}
\epsfxsize=12cm
\centerline{\epsfbox{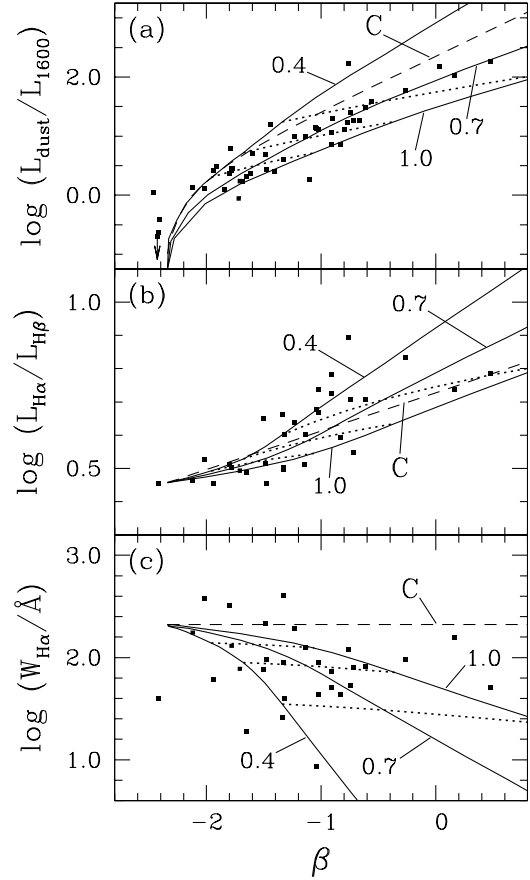}}
\figcaption[f6.eps]{\small Sequences of models with different dust content and
$\hat{\tau}_V^{\rm BC}=2\hat{\tau}_V^{\rm ISM}$. The solid curves correspond to 
different exponents $n$ in the relation $\hat{\tau}_\lambda^{\rm ISM}\propto\lambda^{
-n}$ (as indicated). Dotted lines join models with $\hat{\tau}_V^{\rm BC}=0.5$, 
1.0, and 2.0 (in order of increasing $\beta$) to indicate the scale. The dashed
curve is a sequence of models with $\hat{\tau}_V^{\rm BC}=0.0$ and $\hat{\tau
}_\lambda^{\rm ISM}$ from Calzetti et al. (1994). The data points in ({\it a}),
({\it b}), and ({\it c}) are the same as in Figs.~2, 3, and 4, respectively.}
\end{figure*}
Since each parameter in our model has a specific influence on the
various integrated spectral properties of starburst galaxies, we can
gain insight into the physical origin of the relations defined by 
these observed quantities. An inspection of Figures~2--4 leads to 
the following conclusions. (1) The wide observed range of ultraviolet 
spectral slopes can be most naturally understood as a spread in the
effective absorption optical depth in the ambient ISM, although 
variations in the lifetimes of the birth clouds could also play some
role. It is worth mentioning that the steep ultraviolet spectra
of the galaxies with largest $\beta$ cannot originate from old stars 
because these galaxies also have large $W_{{\rm H}\alpha}$. (2) As noted
above, absorption in the birth clouds is required to explain the large 
H$\alpha/$H$\beta$ ratios observed in galaxies with intermediate ultraviolet
spectral slopes. (3) The fact that $L_{\rm dust}/ L_{1600}$, $L_{{\rm H}
\alpha}/L_{{\rm H} \beta}$, and $\beta$ are all very low at one end
of the observed relations and all very high at the other 
end further suggests that variations in the effective optical depths of
the birth clouds and the ambient ISM may be related (Figs.~2 and 3).
To explore the effects of varying simultaneously $\hat{\tau}_V^{\rm BC}$
and $\hat{\tau }_V^{\rm ISM}$, we assume that they are related by
$\hat{\tau}_V^{\rm BC}= 2 \hat{\tau}_V^{\rm ISM}$ and fix all other
parameters at the standard values given by equation~(\ref{standard}).
The results are shown in Figure~6. Evidently, the models reproduce well
the observed relations between $L_{\rm dust }/L_{1600}$, $L_{{\rm H} 
\alpha}/ L_{{\rm H} \beta}$, and $\beta$ in Figures~6{\it a} and 6{\it b}.
They also fall roughly in the middle of the observed ranges of $W_{{\rm
H}\alpha}$ and $\beta$ in Figure~6{\it c}. 

An important result of our analysis is that the shape of the relation
between $L_{\rm dust}/L_{1600}$ and $\beta$ actually {\it requires}
that the exponent of the effective absorption curve $\hat{
\tau}_\lambda^{\rm ISM} \propto \lambda^{-n}$ be $n\approx0.7$.
In Figure~6, we also show models with $\hat{\tau}_V^{\rm BC}= 2 
\hat{\tau}_V^{\rm ISM}$ as above, but for $n=0.4$ and 1.0 instead of
0.7. These models miss the observed relation, the sequence being too
steep for $n=0.4$ and too shallow for $n=1.0$. This observational 
constraint on $n$ applies essentially to $\hat{\tau }_\lambda^{\rm
ISM}$, since $L_{\rm dust}/L_{1600}$ and $\beta$ are almost 
independent of $\hat{\tau}_\lambda^{\rm BC}$ for $\hat{\tau}_V^{\rm
BC} \ga1$. For $n=0.7$, therefore, the relations defined by $L_{\rm
dust}/L_{1600}$, $L_{{\rm H} \alpha}/ L_{{\rm H} \beta}$, and $\beta$
can be understood most simply as a sequence in the overall 
dust content of galaxies. Our assumption of $\hat{\tau}_V^{\rm BC}=
2\hat{\tau}_V^{\rm ISM}$ should be taken merely as illustrative.
With $n=0.7$, the observations can be matched by a variety of models 
that range from $\hat{\tau}_V^{\rm BC}+\hat{ \tau}_V^{\rm ISM} =0$ at
one end of the relations (with $\beta= -2.3$) to $\hat{\tau}_V^{\rm
BC}\ga1.0$ and $\hat{\tau}_V^{\rm ISM}\approx 1.5$ at the other end
(with $\beta\approx+0.2$). For reference, the fraction of the total
stellar radiation absorbed by dust in the models of Figure~6 ranges
from 0\% to about 90\% along this sequence and is roughly 70\%
for our standard model. Non-ionizing continuum photons account
for about 90\% of the heating in all models except for those with very
low dust content, in which the relative contribution by L$\alpha$ 
photons is more important (over 20\% for $\beta \la-2.2$). As 
Figures~2--4 indicate, the intrinsic scatter about the relations at
fixed $\beta$ can arise from variations in the effective age of the
starburst, the parameters controlling the birth clouds, and the 
fraction of dust in the H{\sc ii} regions (and possibly the IMF and
metallicity of the stars).

\begin{figure*}
\epsfxsize=10.7cm
\centerline{\epsfbox{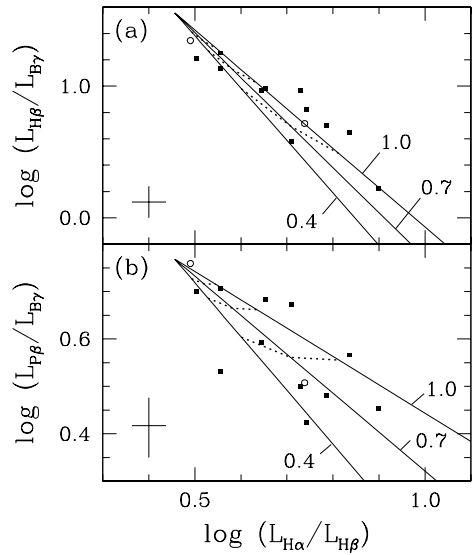}}
\figcaption[f7.eps]{\small ({\it a}) H$\beta$/B$\gamma$ ratio and ({\it b})
P$\beta$/B$\gamma$ ratio plotted against H$\alpha/$H$\beta$ ratio. The data
points are from Calzetti et al. (1996) and include 11 galaxies from the 
Meurer et al. (1999) sample ({\it filled squares}) plus two additional
nearby starburst galaxies ({\it open circles}). The solid and dotted curves
are the same models as in Fig.~6. Typical measurement errors are indicated
in each panel.}
\end{figure*}
Our models are also consistent with observations of near-infrared
H-recombination lines in starburst galaxies. The fluxes of
the Paschen~$\beta$ (1.28~$\mu$m) and Brackett~$\gamma$ (2.17~$\mu$m) 
lines of 11 galaxies in the Meurer et al. (1999) sample are available
from Calzetti et al. (1996). We compute the ${{\rm P }\beta}$ and 
${{\rm B} \gamma}$ luminosities, $L_{{\rm P } \beta}$ and $L_{{\rm B}
\gamma}$, using the same procedure as described in \S2 for
$L_{{\rm H }\alpha}$ and $L_{{\rm H} \beta}$. Figure~7 shows the 
${{\rm H}\beta}/ {{\rm B}\gamma}$ and ${{\rm P } \beta} /{{\rm B} 
\gamma}$ ratios as a function of the ${{\rm H}\alpha} /{{\rm H} 
\beta}$ ratio for the observations and for the same models as in 
Figure~6. Sequences of models are straight lines in these log-log
diagrams, reflecting the assumed power-law form of the effective
absorption curves in the birth clouds and the ambient ISM. Figure~7
shows that $n=0.7$ is consistent with, although not tightly 
constrained by, the observed strengths of optical and near-infrared
recombinations lines in nearby starburst galaxies. As noted above, 
the exponent $n$ of the effective absorption curve is much better
constrained by the relation between the ratio of far-infrared to 
ultraviolet luminosities and the ultraviolet spectral slope.

It is worth noting that the observed mean relations can be 
reproduced by the following simple recipe: use an effective absorption
curve proportional to $\lambda^{-0.7}$ to attenuate the non-ionizing 
continuum radiation and emission lines (as predicted by dust-free
case~B recombination) from each stellar generation, and lower the
normalization of the curve typically by a factor of 3 after $10^7\,$yr
to account for the dispersal of the birth clouds. This recipe
approximates the results of our full calculations shown in 
Figures~6 and 7 but differs from them in that it omits the 
absorption of ionizing photons by dust in H{\sc ii} regions (i.e., 
it assumes $f=0.0$). As mentioned earlier, this has no effect on 
the ultraviolet spectrum and a negligible effect on the far-infrared
luminosity ($<1$\% underestimate) and the ${{\rm H}\alpha}/{{\rm H}
\beta}$ ratio (from $<1$\% to about 7\% overestimate along the 
observed relations). The effect is more important for the H$\alpha$
equivalent width (from 3\% to over 70\% overestimate). Of course, 
the simple recipe above does not allow one to reproduce the scatter
in the observed relations, which requires variations in several 
parameters of our model. However, if only the mean properties are
considered, the simple recipe should suffice for many purposes. 
Both our complete model and the simple recipe derived from it can
be incorporated easily into any population synthesis model.

\begin{figure*}
\epsfxsize=10.7cm
\centerline{\epsfbox{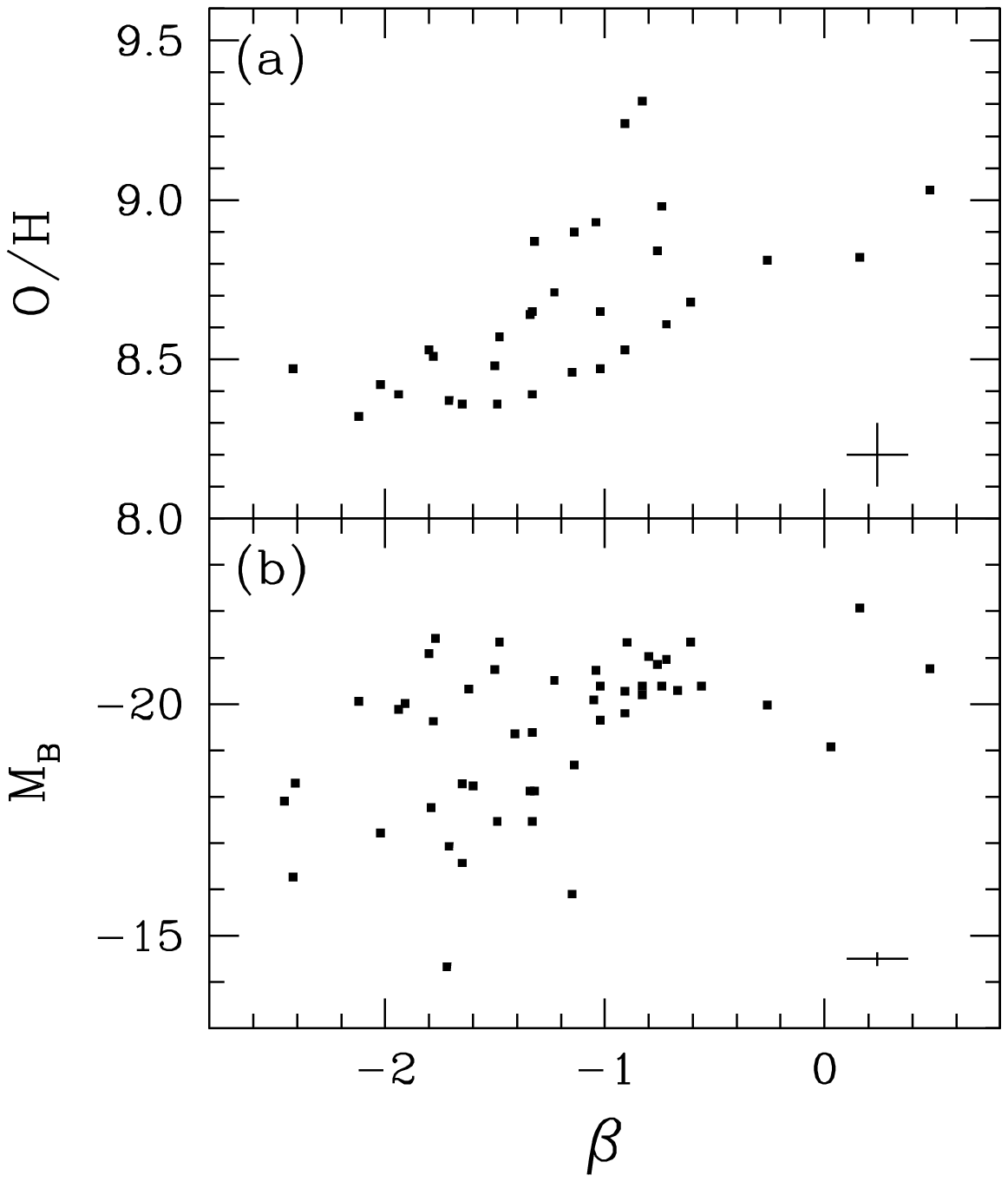}}
\figcaption[f8.eps]{\small ({\it a}) Oxygen abundance plotted against ultraviolet 
spectral slope for galaxies in the Meurer et al. (1999) sample discussed in 
\S3. The abundance measurements are from Calzetti et al. (1994). For
reference, the solar abundance is $({\rm O/H})_\odot\approx8.9$.
({\it b}) Absolute $B$ magnitude plotted against ultraviolet spectral
slope for the galaxies in the Meurer et al. (1999) sample. The adopted Hubble
constant is $H_0= 70$~km$\,$s$^{-1}$Mpc$^{-1}$. Typical measurement errors are
indicated in each panel.}
\end{figure*}
In an attempt to understand the physical basis for the observed 
relation between $L_{\rm dust}/L_{1600}$ and $\beta$, we plot 
other properties of the galaxies against $\beta$ in Figure~8:
the oxygen abundance of H{\sc ii} regions O/H and the absolute $B$ 
magnitude $M_B$. The oxygen abundances of 31 galaxies of the Meurer
et al. (1999) sample, measured in the same aperture as $\beta$, are
available from Calzetti et al. (1994), while the absolute $B$ 
magnitudes of all the galaxies are available from Kinney et al. 
(1993). The Spearman rank correlation coefficients are $r_s= 0.76$
for the relation between O/H and $\beta$ and $r_s= -0.50$ for that
between $M_B$ and $\beta$. For the different numbers of galaxies in
each case, these correlations are significant at the $4 \sigma$ and
$3\sigma$ levels. Thus, they are probably both statistically 
significant, especially the one between O/H and $\beta$ (which was
noted by Calzetti et al. 1994 from a slightly different sample), but
both have very large scatter. This suggests that neither O/H nor 
$M_B$ are entirely responsible for the relations between $L_{\rm dust}/
L_{1600}$, $L_{{\rm H} \alpha}/L_{{\rm H} \beta}$, and $\beta$. The
ultimate physical explanation for these relations therefore requires
further investigation.

For reference, we have also computed models using the Calzetti et al.
(1994) effective absorption curve. This was derived
from the analysis of {\it IUE} spectra of a sample of 33 starburst
galaxies, nearly all of which are included in the Meurer et al. (1999) 
sample. Calzetti et al. (1994) condensed all the spectra in their sample 
into six ``templates'' corresponding to different values of the 
H$\alpha/$H$\beta$ ratio. They then derived an effective absorption
curve by dividing the five spectral templates with the highest $L_{{\rm H}
\alpha}/L_{{\rm H} \beta}$ by the template with the lowest $L_{{\rm H} 
\alpha}/L_{{\rm H} \beta}$ (assuming that the latter represented the
dust-free case). The result is strictly an effective absorption
curve in the sense defined by equation (\ref{taueff}), although it
is often referred to by other names (see footnote~2). Figure~5
shows that the Calzetti et al. (1994) effective absorption curve
is greyer than those derived here. In Figure~6, we show 
the results obtained when adopting the Calzetti et al. (1994) 
curve for $\hat{\tau}_\lambda^{\rm ISM}$ and assuming $\hat{\tau
}_V^{\rm BC}=0$ for consistency.\footnote{To extend the Calzetti et al.
(1994) curve into the near-infrared over the range $0.8\leq \lambda \leq 2.2
\,\micron$, we use the formula provided by Calzetti (1999). To compute 
$\hat{\tau}_\lambda^{\rm ISM}$ at wavelengths $912\leq \lambda\leq 1200
\,${\AA}, we extrapolate the curve linearly in $\log\lambda$. This has 
a negligible influence on the results because typically less than 25\%
of $L_{\rm dust}$ is produced by the absorption blueward of 1200~{\AA}.
In fact, adopting $\hat{\tau}_\lambda^{\rm ISM}\equiv \hat{\tau}_{1200
}^{\rm ISM}$ for $\lambda\leq 1200\,${\AA} gives similar results.} As
anticipated, the models predict a steeper relation between $L_{\rm dust}
/L_{1600}$ and $\beta$ than is observed. The greyness of the Calzetti et
al. (1994) curve probably results from a bias inherent in the method 
used to derive it. As Figure~6{\it b} shows, galaxies with the lowest 
observed H$\alpha/$H$\beta$ ratios extend over a fairly wide range of 
ultraviolet spectral slopes. Thus, the template spectrum with the lowest 
$L_{{\rm H} \alpha}/L_{{\rm H} \beta}$, built by Calzetti et al. 
(1994) from eight spectra with $-2.08\leq \beta \leq -1.33$, may be
relatively but not perfectly dust free, as is assumed by the method.

\section{Comparisons with Observations: the Spatial Distribution of Dust}

Up to now, we have compared our model with observations only in terms
of the effective absorption curve, i.e., without specifying the spatial
distribution and optical properties of the dust. In this section, we
study the influence of different spatial distributions of dust on the 
integrated spectral properties of galaxies. As shown in \S3, the 
effective absorption curve favored by observations of nearby starburst
galaxies is significantly shallower than most known extinction 
curves. This is a natural consequence of any spatial distribution
of dust in which the least extincted stars contribute most to the 
emergent light at short wavelengths. Such an effect has been 
quantified for many distributions of dust, both patchy and smooth
(e.g., Natta \& Panagia 1984; Caplan \& Deharveng 1986; Bruzual, Magris,
\& Calvet 1988; Disney , Davies, \& Phillipps 1989; Witt et al. 1992;
Calzetti et al. 1994, 1996; Puxley \& Brand 1994; Di~Bartolomeo, Barbaro,
\& Perinotto 1995; Bianchi, Ferrara, \& Giovanardi 1996; Witt \&
Gordon 1996; Gordon et al. 1997). These studies, however, did not include
the finite lifetimes of stellar birth clouds.  Thus, they cannot account
for the different attenuation of line and continuum photons in starburst 
galaxies. We now investigate the influence of the spatial distribution
of the dust on the effective absorption curve in the context of our model.

\begin{figure*}
\epsfxsize=8cm
\centerline{\epsfbox{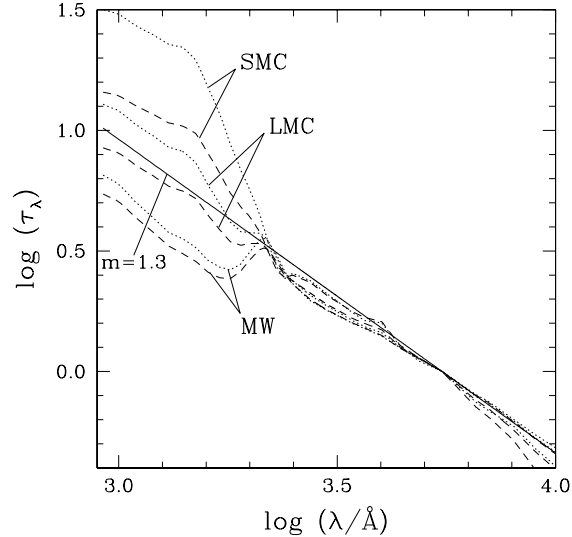}}
\figcaption[f9.eps]{\small Optical depth of graphite-silicates dust plotted
against wavelength. The dotted lines are the true absorption curves of the
Milky Way, LMC, and SMC, while the dashed lines are the corresponding curves
when isotropic scattering is included (eq.~[\ref{tscat}]). These are based
on the Draine \& Lee (1984) model but with the proportions of graphite and
silicates adjusted so as to fit the observed mean extinction curves of the
three galaxies (Pei 1992). The solid line is a power law of index $m=1.3$.
All curves are normalized to unity at 5500~{\AA}.}
\end{figure*}
For simplicity, we assume that the optical depth of the dust has a
power-law form
\begin{equation}
\tau_\lambda\propto\lambda^{-m}\,,
\label{mlam}
\end{equation}
where the choice of $m$ is discussed below. This is consistent
with recent evidence that the optical depth of the dust in nearby 
starburst galaxies does not exhibit a strong feature near 2175~{\AA} 
(Gordon et al. 1997). We further assume for simplicity that 
$\hat{\tau}_\lambda^{\rm BC}$ has the same wavelength dependence
as $\tau_\lambda$. For comparison, Figure~9 shows $\tau_\lambda$ 
plotted against $\lambda$ for Milky Way-type, LMC (30 Doradus)-type,
and SMC-type dust and for the two cases of forward-only and isotropic
scattering (eq.~[\ref{tscat}]). These optical properties are based on
the standard Draine \& Lee (1984) grain model but with the
proportions of graphite and silicates adjusted so as to fit the observed
mean extinction curves of the three galaxies (Pei 1992). The inclusion 
of scattering makes $\tau_\lambda$ greyer because the albedo is larger 
at optical than at ultraviolet wavelengths, the effect increasing from 
Milky Way- to LMC- to SMC-type dust. Power-law fits to these curves 
over the wavelength range $0.1\leq \lambda\leq1\,\micron$ yield values
of $m$ between $1.1$ and $1.5$, with $m=1.3$ representing a middle 
value (Fig.~9). 

The ambient ISM of starburst galaxies cannot be idealized as a 
foreground screen if the dust within them is similar to that in 
the Milky Way, the LMC, or the SMC because the effective 
absorption curve would then take the form $\hat{\tau}_\lambda^{\rm
ISM}\propto\lambda^{-m}$ with $1.1\la m\la1.5$. As shown in
\S3, this would lead to a relation between $L_{\rm dust}
/L_{1600}$ and $\beta$ shallower than observed. We now examine
the other two models of the ambient ISM specified in \S2.
Again, these should be regarded only as convenient 
approximations to more realistic distributions of dust. In
the following, we adopt the same effective starburst age, $t$,
lifetime of birth clouds, $t_{\rm BC}$, and fraction of dust in 
the ionized gas, $f$, as in our standard model in \S3
(eq.~[\ref{standard}]). 

\begin{figure*}
\epsfxsize=8cm
\centerline{\epsfbox{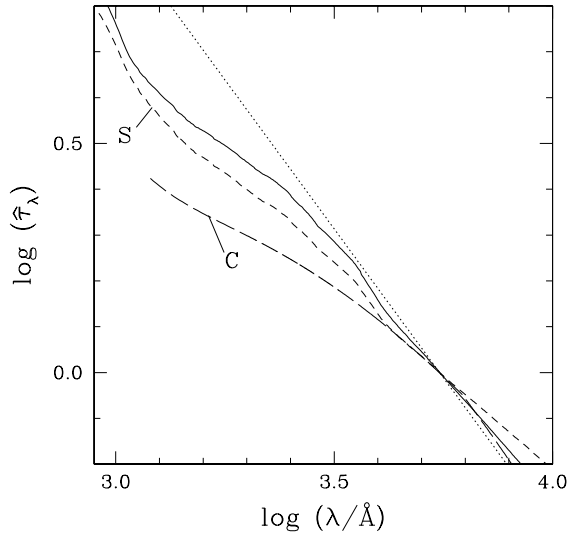}}
\figcaption[f10.eps]{\small Effective absorption curve (as defined by
eq.~[\ref{taueff}]) of a model with effective starburst age $t=3\times10^8\,$yr
in which the ambient ISM is represented by a mixed slab ({\it solid line}).
The optical depth of the dust is assumed to have the form $\tau_\lambda
\propto\lambda^{-m}$ with $m=1.3$ ({\it dotted line}). The short-dashed curve
is the standard model of eq.~(\ref{standard}), and the long-dashed curve is
from Calzetti et al. (1994). All curves are normalized to unity at 5500~{\AA}.}
\end{figure*}
We first consider the mixed slab model for the ambient ISM. In this case,
$\hat{\tau}_\lambda^{\rm ISM}$ is related to the optical depth through the
slab in the direction normal to the surface, $\tau_\lambda^{ \rm sl}
\propto\lambda^{-m}$, by equations~(\ref{tism}) and (\ref{tslab}). By
analogy with our standard model in \S3, we adopt $\hat{\tau}_V^{\rm BC}=1.0$
for the optical depth of the birth clouds and take $\tau_V^{\rm sl}=0.5$.
Figure~10 shows the total effective absorption curve at $t=3\times
10^8\,$yr with $m=1.3$. A comparison with Figure~5 shows 
that this is very similar to the total effective absorption curve of the
standard model with $n=0.7$ in \S3. To understand the origin of this
result, we plotted $\hat{\tau}_\lambda^{\rm ISM}$ as a function of 
$\tau_\lambda^{\rm sl}$ and noticed that the simple formula $\hat{\tau
}_\lambda^{\rm ISM}\approx({ \tau_\lambda^{\rm sl}})^{1/2}$ approximates
$\hat{\tau}_\lambda^{\rm ISM}$ with an accuracy or 5\% or better over the
whole range $1\la \tau_\lambda^{\rm sl}\la10$ (at low optical depths,
$\hat{\tau}_\lambda^{\rm ISM}$ is closer to $\tau_\lambda^{\rm sl}$). This
weak dependence of $\hat{\tau}_\lambda^{\rm ISM}$ on $\tau_\lambda^{\rm
sl}$ arises from the fact that, for large $\tau_\lambda^{\rm sl}$, sources 
located at high optical depths from the surface of the slab contribute very
little to the emergent light. In the ultraviolet, therefore, where
$\tau_\lambda^{\rm sl}$ is greater than unity, the wavelength dependence
of $\hat{\tau }_\lambda^{\rm ISM}$ is closer to $\lambda^{-m/2}$ than to 
$\lambda^{-m}$ in the model of Figure~10. Hence, for reasonable assumptions
about the optical properties of the dust, a mixed slab model for the ambient
ISM can produce an effective absorption curve similar to that required
by the observed relation between $L_{\rm dust}/L_{1600}$ and $\beta$.

\begin{figure*}
\epsfxsize=12cm
\centerline{\epsfbox{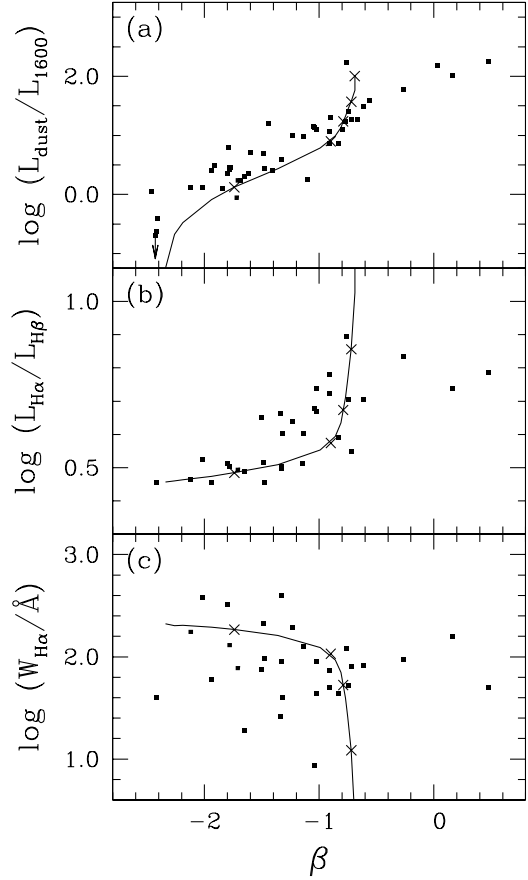}}
\figcaption[f11.eps]{\small Sequence of models in which the ambient ISM is
represented by a mixed slab with different dust content and $\hat{\tau}_V^{\rm
BC}=2\tau_V^{\rm sl}$.  The optical depth of the dust is assumed to have the
form $\tau_\lambda\propto \lambda^{-m}$ with $m=1.3$. Crosses mark the models
with $\hat{\tau}_V^{\rm BC}=0.1$, 0.5, 1.0, 2.0, and 5.0 (in order of increasing
$\beta$) to indicate the scale.  The data points in ({\it a}), ({\it b}), and
({\it c}) are the same as in Figs.~2, 3, and 4, respectively.}
\end{figure*}
Mixed slab models, however, cannot accommodate the full range of
observations. In Figure~11, we show the analog of Figure~6 for models
with $\hat{\tau}_V^{\rm BC}=2 \tau_V^{\rm sl}$ and $m=1.3$. At low optical 
depths, i.e. for $\hat{\tau}_V^{\rm BC}\la1$, the models reproduce well 
the observed relation between $L_{\rm dust}/L_{1600}$ and $\beta$, as 
noted above. At larger optical depths, however, the models deviate 
markedly from the observations. This upturn, which persists over
the range $1.0 \leq m\leq1.5$, is entirely a consequence of 
absorption in the ambient ISM because, for large optical depths of 
the birth clouds, i.e. for $\hat{\tau}_V^{\rm BC} \ga 1$, $\beta$ and
$L_{\rm dust}/L_{1600}$ are controlled by $\hat{\tau}_\lambda^{\rm ISM}$
(Figs.~2{\it c} and 2{\it d}). The reason for this behavior can be traced 
back to the fact that $\hat{\tau }_\lambda^{\rm ISM}$ scales only as 
$(\tau_\lambda^{\rm sl})^{1/2}$ for $\tau_\lambda^{\rm sl} \ga 1$.
As $\tau_V^{\rm sl}$ increases in Figure~11, $\hat{\tau}_\lambda^{\rm
ISM}$ approaches a power law of index $m/2$ over a range that
expands toward progressively longer wavelengths (since $\tau_\lambda^{\rm 
sl}\propto \tau_V^{ \rm sl}\lambda^{ -m}$). Once the power-law shape
extends over the entire ultraviolet wavelength range, increasing 
$\tau_V^{\rm sl}$ no longer affects $\beta$ but continues to alter
$L_{\rm dust}/L_{1600}$, $L_{{\rm H}\alpha} /L_{{\rm H} \beta}$, and
$W_{{\rm H} \alpha}$. This is an intrinsic limitation of the mixed slab
model. For large $\tau_V^{\rm sl}$, $\beta$ cannot reach the very 
large values observed because sources at high optical depths from the
surface of the slab contribute very little to the emergent ultraviolet
radiation.

\begin{figure*}
\epsfxsize=8cm
\centerline{\epsfbox{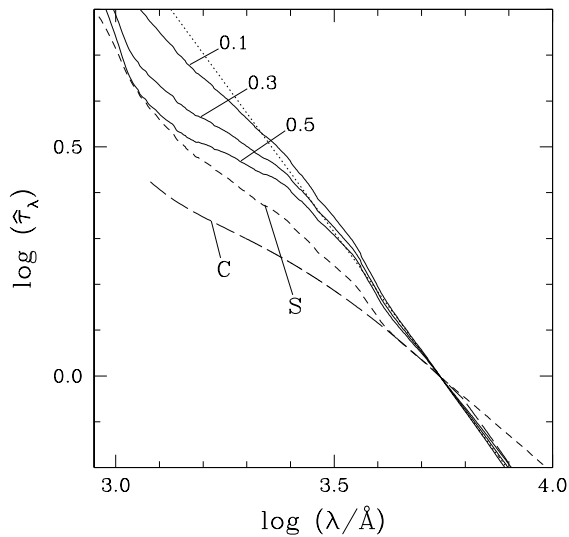}}
\figcaption[f12.eps]{\small Effective absorption curves (as defined by
eq.~[\ref{taueff}]) of models with effective starburst age $t=3\times10^8\,$yr
in which the ambient ISM is represented by a Poisson distribution of clouds.
The optical depth of the dust is assumed to have the form
$\tau_\lambda\propto\lambda^{-m}$ with $m=1.3$ ({\it dotted line}). The
models are shown for different cloud optical depths, $\tau_V^c=0.1$, 0.3, and
0.5 (as indicated). The short-dashed curve is the standard model of
eq.~(\ref{standard}), and the long-dashed curve is from Calzetti et al.
(1994). All curves are normalized to unity at 5500~{\AA}.}
\end{figure*}
We now consider the discrete cloud model for the ambient
ISM. In this case, $\hat{\tau}_\lambda^{\rm ISM}$ is related to
the optical depth per cloud, $\tau_\lambda^c\propto\lambda^{-m}$,
and the mean number of clouds along a line of sight, $\overline{n}$, by
the expression $\hat{\tau}_\lambda^{\rm ISM}=\overline{n}\,\left[1-\exp(- 
\tau_\lambda^c) \right]$ (eqs.~[\ref{tism}] and [\ref{tpois}]). Thus,
$\overline{n}$ does not affect the wavelength dependence of 
$\hat{\tau}_\lambda^{\rm ISM}$. To illustrate the effective absorption 
curves of the discrete cloud models, we fix $\hat{\tau}_V^{\rm BC}=1.0$
and experiment with different optical depths of the clouds, 
$\tau_V^c= 0.1$, 0.3, and 0.5. By analogy with our standard model in 
\S3, we adjust $\overline{n}$ so that the mean optical depth along a 
line of sight is $\overline{n}\tau_V^c =0.5$ in each case. Figure~12
shows the total effective absorption curves at $t=3\times10^8\,$yr
for these models with $m=1.3$. The models exhibit the characteristic 
signatures of the absorption in the birth clouds (Fig.~5). As
$\tau_V^c$ increases, $\hat{\tau}_\lambda^{\rm }$ becomes greyer
because the wavelength below which the clouds absorb all incident
radiation, i.e. where $\hat{\tau}_\lambda^{\rm ISM} \approx\overline{n}$,
becomes progressively larger. Figure~12 indicates that the model
with $\tau_V^c =0.3$ has the ultraviolet slope closest to that
of the standard model in \S3. For reasonable assumptions about the
optical properties of the dust, therefore, discrete cloud models also
lead to effective absorption curves similar to that required by
the observed relation between $L_{\rm dust}/L_{1600}$ and $\beta$.

\begin{figure*}
\epsfxsize=12cm
\centerline{\epsfbox{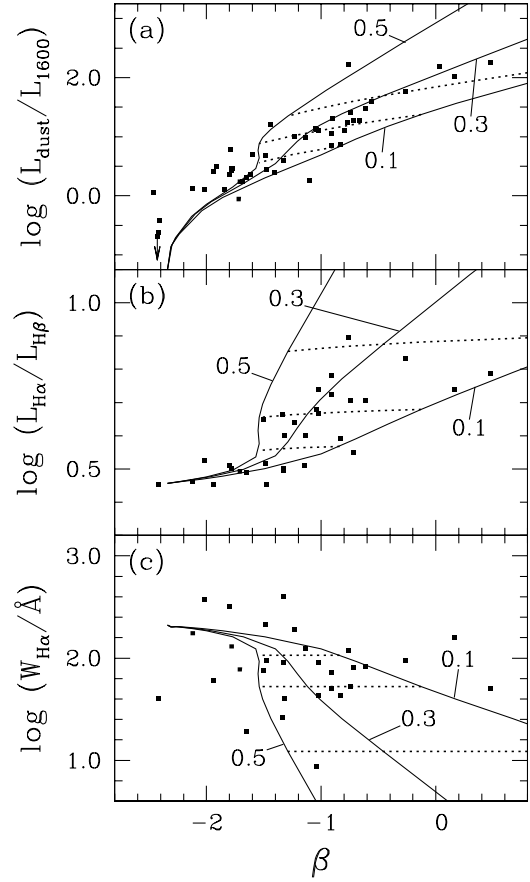}}
\figcaption[f13.eps]{\small Sequences of models in which the ambient ISM is
represented by a Poisson distribution of clouds with different dust content and
$\hat{\tau}_V^{\rm BC}=2 \overline{n}\tau_V^c$. The optical depth of the dust
is assumed to have the form $\tau_\lambda\propto \lambda^{-m}$ with $m=1.3$.
The models are shown for different cloud optical depths, $\tau_V^c =0.1$, 0.3,
and 0.5 (as indicated). Dotted lines join models with $\hat{\tau}_V^{\rm BC}=
0.5$, 1.0, and 2.0 (in order of increasing $\beta$) to indicate the scale. The
data points in ({\it a}), ({\it b}), and ({\it c}) are the same as in Figs.~2,
3, and 4, respectively.}
\end{figure*}
The advantage of discrete cloud models over mixed slab models
is that they can account for the full range of observations.
In Figure~13, we show the analog of Figure~6 for models with 
$\hat{\tau}_V^{\rm BC}=2\overline{n} \tau_V^c$, $\overline{n}$ 
a continuous variable, $\tau_V^c= 0.1$, 0.3, and 0.5, and $m=1.3$.
The models with $\tau_V^c =0.3$ agree with all the observations. 
In particular, they match remarkably well the relations between 
$L_{\rm dust}/ L_{1600}$, $L_{{\rm H}\alpha} /L_{{\rm H} \beta}$,
and $\beta$ over the entire range observed. Models with $\tau_V^c
=0.5$ and 0.1 lie above and below those with $\tau_V^c =0.3$, as 
expected from the different effective absorption curves in Figure~12.
The mean number of clouds in the models with $\tau_V^c= 0.3$ ranges
from $\overline{n}=0$ at one end of the relations (with $\beta=-2.3$)
to $\overline{n} \approx5$ at the other end (with $\beta\approx+
0.2$). These values of $\overline{n}$ are small enough that
the distribution of dust can be regarded as patchy over most
of the range. We also find that the relations between $L_{\rm dust}
/ L_{1600}$, $L_{{\rm H}\alpha} / L_{{\rm H}\beta}$, and $\beta$ 
are very stable to $m$ over the range $1.0 \leq m\leq1.5$ for 
$\tau_V^c = 0.3$, but curiously less stable for $\tau_V^c = 0.1$
and 0.5. The above values of $\tau_V^c$ would be somewhat smaller
if we had adopted an optical depth $\tau_\lambda$ with a strong
feature near 2175~{\AA}.

Our results demonstrate that, for reasonable assumptions about
the wavelength dependence of the dust optical depth $\tau_\lambda$,
discrete cloud models can account for all the observed 
integrated spectral properties of nearby starburst galaxies.
Poisson distributions of clouds have traditionally been used to 
interpret variations in the observed reddening of stars at the 
same distance within the Milky Way. These observations can be well
represented by a model with two types of clouds, one with 
$A_V^c \approx0.2$ and another, less abundant one with $A_V^c 
\approx 0.9$ (Spitzer 1978). This is similar to our model, which
has a population of optically thin clouds in the ambient ISM and 
a population of optically thick birth clouds. It is worth 
emphasizing that in any model with discrete clouds, the optical 
depth has a large variance along different lines of sight within
a given galaxy, depending on the number of clouds encountered. In
particular, some lines of sight can be quite opaque, even for
$\tau_V^c = 0.3$ and $\overline{n}\sim1$. For starburst galaxies
with the largest values of $L_{\rm dust}/L_{1600}$, $\overline{n}
\approx5$ is indicated, and most lines of sight are opaque. Thus, there is 
no contradiction between the relatively small values of $\tau_V^c$
found here and the very dusty appearance of some starburst galaxies.

\section{Conclusions}

We have developed a simple model to compute the effects of dust
on the integrated spectral properties of galaxies, based on 
an idealized prescription of the main features of the ISM. Our
model includes the ionization of H{\sc ii} regions
in the interiors of the dense clouds in which stars form. Emission
lines from H{\sc ii} regions and the non-ionizing continuum from
young stars are attenuated in the same way by dust in the outer H{\sc
i} envelopes of the birth clouds and the ambient ISM. However, since
the model also includes the finite lifetimes of the birth clouds,
the non-ionizing continuum radiation from stars that live longer
than the birth clouds is attenuated only by the ambient ISM. 
We show that this can fully resolve the apparent discrepancy
between the attenuation of line and continuum photons in starburst 
galaxies. This enables us, in turn, to interpret in a consistent
way all the observations of a homogeneous sample of nearby 
ultraviolet-selected starburst galaxies, including the
ratio of far-infrared to ultraviolet
luminosities ($L_{\rm dust}/L_{1600}$), the ratio of H$\alpha$ to
H$\beta$ luminosities ($L_{{\rm H}\alpha}/L_{{\rm H}\beta}$), the 
H$\alpha$ equivalent width ($W_{{\rm H}\alpha}$), and the 
ultraviolet spectral slope ($\beta$). 

The different parameters in our model, including the 
effective age of the starburst, the lifetime and effective 
optical depth of the stellar birth clouds, the effective optical
depth in the ambient ISM, and the fraction of dust in the ionized
gas, each have a specific influence on the integrated spectral
properties $L_{\rm dust} /L_{1600}$, $L_{{\rm H}\alpha}/L_{{\rm
H}\beta}$, $W_{{\rm H} \alpha}$, and $\beta$. This provides
new insights into the origin of the mean relations 
defined by the data and the scatter about these relations.
In particular, the relation between the ratio of far-infrared
to ultraviolet luminosities and the ultraviolet spectral slope
in starburst galaxies reflects the wavelength dependence of the
effective absorption in the ambient ISM. We find that a power 
law of the form $\hat{\tau }_\lambda^{\rm ISM} \propto 
\lambda^{ -0.7}$ accounts remarkably well for all the 
observations. The relation between $L_{\rm dust} /L_{1600}$
and $\beta$ can then be interpreted as a sequence in
the overall dust content of the galaxies. Interestingly, this
relation is accompanied by much weaker trends of the oxygen 
abundance and the optical luminosity with the ultraviolet 
spectral slope. The fact that our model reproduces the observed
spectral properties of nearby starburst galaxies relatively
easily leads us to suspect that, with suitable adjustment of the
parameters, it could also reproduce those of more quiescent 
(but still star-forming) galaxies.

The effective absorption curve required by the observations
is much greyer than would be produced by a foreground screen
of dust like that in the Milky Way, the LMC, or the SMC. We
have explored whether this could be accounted for by a
much steeper wavelength dependence of the optical depth, i.e. 
$\tau_\lambda \propto \lambda^{-m}$ with $1.0\la m \la 1.5$,
combined with a more realistic spatial distribution of the 
dust. We find that a mixed slab model for the ambient ISM can
produce the required effective absorption curve for low dust
content but cannot explain the observations of starburst 
galaxies with very reddened ultraviolet spectra. In contrast,
we show that a random distribution of discrete clouds
provides a consistent interpretation of all the observed 
integrated spectral properties of starburst galaxies. While
these results were anticipated in some previous studies,
we have shown here for the first time how to reconcile them
with the large H$\alpha/$H$\beta$ ratios and other observations.
We also find that the optical depths of the clouds 
favored by our analysis are similar to those inferred from
the statistics of stellar reddening in the Milky Way.

The model we have developed for computing the absorption of 
starlight by dust in galaxies can be combined easily, by design,
with any population synthesis model. The observed mean 
relations for starburst galaxies can also be reproduced by 
the following simple recipe: use an effective absorption curve 
proportional to $\lambda^{-0.7}$ to attenuate the line and 
continuum radiation from each stellar generation, and lower the
normalization of the curve typically by a factor of 3 after 
$10^7\,$yr to account for the dispersal of the birth clouds.
This recipe accounts at least as well as the one by Calzetti et
al. (1994, and as modified by Calzetti 1997, 1999) for the 
effects of dust on the non-ionizing continuum radiation. In 
addition, it fully resolves the apparent discrepancy
between the attenuation of line and continuum photons in 
starburst galaxies. We believe, therefore, that our 
model and the recipe derived from it provide simple yet 
versatile tools to interpret the integrated spectral 
properties of starburst and possibly other types of 
galaxies. In future work, we plan to apply them to the 
growing body of observations of high-redshift galaxies.

\acknowledgements

We thank P.~Boiss\'e, A.~Ferrara, and N.~Panagia for valuable 
discussions. S.C. appreciates the hospitality of the STScI, and
S.M.F. that of the IAP, during the course of several visits. 
This research was supported in part by the National Science 
Foundation through grant no.~PHY94-07194 to the Institute for 
Theoretical Physics.

\end{document}